\documentclass[twocolumn,times]{aastex63}

\usepackage{amsmath}

\received{2020 September 29}
\revised{2021 May 14}
\accepted{2021 May 30}
\submitjournal{The Astrophysical Journal}

\shorttitle{Outflows in LERGs}
\shortauthors{Singha et al.}

\graphicspath{{./}{figures/}}

\begin{document}

\title{Ionized Gas Outflows in Low Excitation Radio Galaxies Are Radiation Driven}

\author[0000-0001-5687-1516]{M.~Singha}
\affiliation{Department of Physics \& Astronomy, University of Manitoba, 30A Sifton Road, Winnipeg, MB R3T 2N2, Canada}

\author[0000-0001-6421-054X]{C.~P.~O'Dea}
\affiliation{Department of Physics \& Astronomy, University of Manitoba, 30A Sifton Road, Winnipeg, MB R3T 2N2, Canada}

\author[0000-0003-1432-253X]{Y.~A.~Gordon}
\affiliation{Department of Physics \& Astronomy, University of Manitoba, 30A Sifton Road, Winnipeg, MB R3T 2N2, Canada}

\author[0000-0002-2958-0593]{C.~Lawlor-Forsyth}
\affiliation{Department of Physics and Astronomy, University of Waterloo, Waterloo, ON N2L 3G1, Canada}
\affiliation{Waterloo Centre for Astrophysics, University of Waterloo, Waterloo, ON N2L 3G1, Canada}
\affiliation{Department of Physics \& Astronomy, University of Manitoba, 30A Sifton Road, Winnipeg, MB R3T 2N2, Canada}

\author[0000-0002-4735-8224 ]{S.~A.~Baum}
\affiliation{Department of Physics \& Astronomy, University of Manitoba, 30A Sifton Road, Winnipeg, MB R3T 2N2, Canada}

\correspondingauthor{M.~Singha}
\email{singham@myumanitoba.ca}

\begin{abstract}

Low excitation radio galaxies (LERGs) are weakly accreting active galactic nuclei (AGN) believed to be fuelled by radiatively inefficient accretion processes. Despite this, recent works have shown evidence for ionized and neutral hydrogen gas outflows in these galaxies. To investigate the potential drivers of such outflows we select a sample of 802 LERGs using the \citet{best2012} catalogue of radio galaxies. By modelling the [\ion{O}{3}]~$\lambda 5007$ profile in Sloan Digital Sky Survey spectra of a sample of 802 LERGs, we determine that the ionized outflows are present in $\sim 1.5\%$ of the population. Using $1.4~\text{GHz}$ imaging from the Faint Images of the Radio Sky at Twenty Centimeters survey we analyze the radio morphology of LERGs with outflows and find these to be consistent with the parent LERG population. However, we note that unlike the majority of the LERG population, those LERGs showing outflows have Eddington scaled accretion rates close to $1\%$. This is indicative that ionized outflows in LERGs are driven by the radiation pressure from the accretion disk of the AGN rather than the radio jets. We report specific star formation rates in the range of $10^{-12} < \text{sSFR} < 10^{-9}~\text{yr}^{-1}$. Moreover, we observe higher mass outflow rates of $7-150~M_{\sun}~\text{yr}^{-1}$ for these LERGs than luminous quasars for a given bolometric luminosity, which could possibly be due to the radio source in LERGs boosting the mass-loading. This scenario could indicate that these outflows could potentially drive feedback in LERGs.

\end{abstract}


\section{Introduction}\label{sec:intro}

Many massive galaxies host supermassive black holes (SMBH) at their centers. These SMBH  grow through accreting gas and are commonly visible as Active Galactic Nuclei (AGN) at the centres of their host galaxies. A fascinating discovery in modern astronomy is that the masses of these SMBHs are proportional to that of the stellar velocity dispersion of their host galaxies \citep[$M$-$\sigma$ relation, e.g.,][]{kormendy1995,magorrian1998,tremaine2002,gultekin2009,kormendy2013}. Contemporaneously, research explaining the bright end of the galaxy luminosity function \citep{bower2006} gave rise to the idea that the AGN is somehow affecting the host galaxy's evolution. In order to explain these results, theoretical models of galaxy formation and evolution came up with the idea of ``AGN feedback,'' during the winds, jet and radiation from the AGN can expel and heat up the gas in their host galaxies, suppressing the gas accretion onto the SMBH and shutting down the star formation \citep{fabian2012}; and this could reproduce the observable properties of the intergalactic and intracluster medium and the massive galaxies such as $M$ -$\sigma$ relations; the sharp cut-off at the bright end of the galaxy luminosity function 
\citep[e.g.,][]{silk1998,churazov2005,bower2006,hopkins2006,mccarthy2010,gaspari2011,harrison2014}. A key area of ongoing research is to place observational constraints on how AGN activity could couple to gas in galaxies and haloes, and the region where the AGN activities are prevalent \citep[see, e.g.,][]{cattaneo2009,alexander2012,fabian2012,mcnamara2012}.

Many successful galaxy evolution models suggest that the AGN could inject an enormous amount of energy into the host galaxy in the form of radiation  (`radiation driven' or `quasar mode feedback'), where AGN could drive large scale (i.e. $0.1-\text{several kpc}$) energetic outflows, and the wind generated close to the AGN flowing through the galaxy pushes the gas out of their host galaxies and as a consequence, it shuts down the star formation and future BH growth; and further enriches the intergalactic medium with metals \citep[e.g.,][]{silk1998,fabian1999,benson2003,hopkins2006,fabian2012}. While there is concrete observational evidence that star formation processes such as supernovae and stellar-winds could potentially drive galactic-scale outflows \citep[e.g.,][]{heckman1990,moorwood1996,lehnert1996,dahlem1997,swinbank2009} which is one of the main mechanisms behind driving galaxy evolution \citep[e.g.,][]{hopkins2006,dallavecchia2008}, it is commonly assumed that only AGN activity could result in outflows with velocities $\sim$ a few 1000 $\text{km s}^{-1}$ and is a crucial ingredient for most massive galaxies' evolution \citep[e.g.,][]{benson2003,mccarthy2011}. AGN with high accretion rates (Eddington ratio, $\lambda_{\text{Edd}}>0.01$) are thought to be radiatively efficient. They typically have thin, bright accretion disks \citep{shakura1973,malkan1983,blaes2007,best2012}. A significant portion of these AGN are radio-loud AGN, with extended radio jets \citep[$\sim$ 10-a few 100 kpc][]{miley1980,blundell1999,hardcastle2019}. Aside from the radiation pressure from the accretion disk, the outflows may well be driven by the acceleration caused by the radio jet - ambient medium interaction \citep[e.g.,][]{morganti2003b,morganti2005b,emonts2005,labiano2013,schulz2018}. 
Numerous studies \citep[e.g.,][]{capetti1999,tadhunter2001} claim that the ambient gas could be swept up by the radio jet and compressed due the radio jets hollowing out a cocoon-like structure on its path. \citet{villarmartin1999} mentioned that high outflow velocities ($>1000~\text{km s}^{-1}$) could occur if the gas clouds are entrained in hot and shocked gas, expanding out behind the bow-shock of the jet \citep[e.g.,][]{stone1992,dai1994,klein1994,odea2002}. \citet{emonts2005} concluded that prevalence of large amounts of neutral gas outflowing could be due to the energy-driven mechanism for the gas outflows.

Contrary to the 'quasar mode feedback', the energy released by the AGN as hot-plasma jets could control the cooling level of the hot gas in many of the massive haloes \citep[`radio mode feedback', see][]{bower2012,harrison2014}. In this feedback mode, the fuelling of the accreted material onto the SMBH produces relativistic plasma jets \citep{best2012}. It is widely believed that the radiatively inefficient ($\lambda_{\text{Edd}} < 0.01$) AGN accretion flows (advection-dominated accretion flows, or ADAFs), which are optically thin, geometrically thick \citep{narayan1995}, tend to emit most of their energy in form of radio jets \citep{merloni2007}.

The current understanding is that the AGN in low excitation radio galaxies (LERGs) drive galaxy evolution mainly via heating up the ambient gas by launching hot plasma jets \citep{best2012,mingo2016}, while the high excitation radio galaxies (HERGs) could drive galaxy evolution by expelling gas out of the host galaxies through outflows \citep[e.g.,][]{morganti2005a,gupta2006,couto2017}, or by launching hot plasma jets and heating up the ambient gas \citep[e.g.,][]{fabian2003,wilson2006,gendron2017}. The situation is further complicated as the photoionization from AGN radiation highly ionizes the surrounding gas along the jet's axis in HERGs \citep{couto2017}. The ambient gas could further be accelerated by radiation pressure and possibly be heated by radiative heating \citep{liu2013}. Classically, LERGs and HERGs have been divided based on the relative intensity of high and low excitation lines in their optical spectra \citep[e.g.,][]{hine1979,laing1994,buttiglione2010}, and are assumed to infer different classes of objects \citep[e.g.,][]{best2012,miraghaei2017}. HERGs tend to be highly accreting ($\lambda_{\text{Edd}} > 0.01$), whereas LERGs show low accretion rates \citep[$\lambda_{\text{Edd}}<0.01$; e.g.,][]{heckman2014,yuan2014}. It has been widely assumed that the accretion mode in HERGs tend to be radiatively efficient, whereas LERGs are mostly radiatively inefficient \citep{best2012}.

There are  fundamental differences between the radiatively inefficient and efficient AGN. \citet{hardcastle2007} has argued that this difference could relate to the origin of the accreting material where the accreted cold gas leads to a stable accretion disk. \citet{yuan2014} argued that the hot accretion flows occur at lower mass accretion rates, and could be described by advection-dominated accretion flow (ADAF) \citep[e.g.;][]{narayan1995,blandford1999}. \citet{narayan1995,blandford1999,yuan2014} showed that these hot accretion flows could possibly lead to strong outflows and jets for these radiatively inefficient AGN. \citet{mcnamara2011,martinez2011,blandford2019} have argued that black hole spin could play an important role in launching the jets. \citet{narayan1995} stated that radio jets could be driven, when the SMBH is accreting at a rate comparable to the Eddington limit or greater than the Eddington-limit. In this scenario, the high optical depth causes the viscous time to be larger than the diffusion time. As a result, the gas cannot cool and advects the dissipated energy.
Recent studies by \citep{kollmeier2006,trump2009,trump2011} have supported the findings of \citet{narayan1995}, finding that very few broad-line AGN would accrete below the Eddington ratio, $\lambda_\text{Edd} \sim 0.01$. \citet{wu2011,ghisellini2011} indicated that a transition between flat-spectrum radio quasars and BL~Lac objects (beamed radiatively-inefficient AGN) could occur when the SMBH accretes around 1\% of its Eddington-rate. The synthesis models describing AGN evolution \citep[e.g.][]{merloni2008,ananna2019} are built upon the two different AGN acrretion modes.

Despite the difficulties in observing gaseous outflows in galaxies that are intrinsically weak-lined, there is some observational evidence of outflows in LERGs. \citet{morganti2003b} detected an outflow in \ion{H}{1} in the LERG 3C~293. \citet{emonts2005} further reported an asymmetric [\ion{S}{2}]~$\lambda\lambda 6717, 6731$ line-profile (``blue-wing''), indicating an ionized outflow ejecting out low-density gas ($300~\text{cm}^{-3}$) in 3C~293. \citet{emonts2005} detected similar kinematics in the ionized gas ([\ion{S}{2}]) compared to the \ion{H}{1}, concluding that the acceleration from the bow shock of the radio jet is driving the outflows. \citet{morganti2005b} further reported an outflow in \ion{H}{1} in the LERG 3C~236, where \citet{labiano2013} also saw signatures of asymmetry in the [\ion{O}{3}] line profile. They stated that the AGN activity could be triggered due to a recent merger.

\citet{chandola2017} studied a sample of 91 radio galaxies at low redshift ($0.02 < z < 0.23$) with low radio luminosity ($10^{23} < L_{1.4~\text{GHz}} < 10^{26}~\text{W Hz}^{-1}$), among which 80 are LERGs. They investigated the absorption profiles of \ion{H}{1} to look for possible feedback by radio jets. They found that the most blue-shifted gas cloud has a velocity shift with respect to the optical systemic velocity, $\sim -310~\text{km s}^{-1}$. Additionally, these 80 LERGs show large line-widths (full width at $20\%$ of the peak) of $\sim 500~\text{km s}^{-1}$, which are associated with radio sources with higher $L_{1.4~\text{GHz}}$. This gas could possibly be disturbed by the radio jet causing the large line-width and shock accelerated in their host galaxies causing this blue-shift in \ion{H}{1}. \citet{chandola2017} concluded that LERGs with relatively higher radio luminosities show evidence of outflows interacting with the interstellar medium which may affect the star formation rates as the sources evolve. Recently, another study by \citet{schulz2018} with very-long-baseline interferometry of 3C~236 suggested that the outflow could be driven by a jet as the radiation will be inefficient to drive the outflows in LERGs.

The broad and asymmetric [\ion{O}{3}]~$\lambda 5007$ line profile has been widely used as a diagnostic tool to trace outflows which could be over $>$ 10 pc scales \citep[e.g.,][]{mullaney2013,harrison2014}. As [\ion{O}{3}] is a forbidden transition line, it cannot be produced in the highly dense subparsec scales of the BLR (broad emission line region), which makes it a good tracer of the NLR (narrow line region) gas kinematics, and it can be observed from 10 pc - several kpc scales \citep[e.g.][]{wampler1975,boroson1985,stockton1987}.

\citet{labiano2013} reported remarkably large widths in \ion{H}{1} and the optical highly-ionized emission lines of [\ion{O}{3}]~$\lambda\lambda 4959, 5007$ in the SDSS spectrum \citep[Sloan Digital Sky Survey,][]{york2000} of LERG 3C~236 where they estimated the $\text{FWHM}$ (full width at half maximum)$\sim 900~\text{km s}^{-1}$, and $\text{FWZI}$ (full width at zero intensity) $\sim 2000~\text{km s}^{-1}$. They found that this large $\text{FWHM}$ and $\text{FWZI}$ were also seen in the optical emission of H$\alpha$, [\ion{N}{2}]~$\lambda\lambda 6548, 6583$ and [\ion{S}{2}]~$\lambda\lambda 6716, 6731$ doublets. Previously, studies by \citet{dasyra2011,guillard2012} reported such large line-widths in the infrared lines of [\ion{Ne}{2}] and [\ion{Ne}{3}]. 
An overlay of the [\ion{O}{3}]~$\lambda 5007$ line and a Gaussian fit to the $^{12}\text{CO~(2--1)}$ line revealed the existence of extreme red (red wing) and blue velocities (blue wing) $\sim 400~\text{km s}^{-1}$, which is far beyond the range allowed by the rotation of the disk \citep{labiano2013}. 
The authors found that the FWZI value of the [\ion{O}{3}]~$\lambda\lambda 4959, 5007$ $\sim 3$ times the FWZI value estimated for the $^{12}\text{CO~(2--1)}$ line. Furthermore, they reported that the FWZI values are similar for the red and blue wings, suggesting that that the red wing could originate in the receding side of the same outflow system. \citet{labiano2013} concluded that these extreme red and blue velocities and the large line width clearly indicates the presence of an outflow. 
\citet{labiano2013} further reported comparing the outflows between [\ion{O}{3}]~$\lambda\lambda 4959, 5007$ and \ion{H}{1} shows that the blue wing from the highly ionized gas cover similar velocity range. This suggests that the outflowing ionized gas may have started to recombine and form atomic \ion{H}{1} \citep{morganti2003a}. One drawback of the study by \citet{labiano2013} was that they lacked any physical spectral modelling of the [\ion{O}{3}]~$\lambda\lambda 4959, 5007$ line and therefore, their measured red and blue velocities may not represent the exact velocity offset of the respective ionized gas cloud from the rest-frame of the host galaxy.

A spectroscopic study of optical [\ion{O}{3}] lines in $\sim$ 39,000 type~2 AGN by \citet{woo2016} reported that the outflows are strongly correlated with the [\ion{O}{3}] luminosity \citep[which is related to the AGN bolometric luminosity as per][]{heckman2014} and the Eddington ratio. The AGN in the sample of \citet{woo2016} consisted of both low-ionization nuclear emission-line regions (LINERs) and Seyferts, which imply that the connection of the AGN accretion disk to the outflows could possibly be applicable to LINERs too. There is a possibility that a sub-population of LINERs could potentially represent a radio quiet analogue of LERGs, but it is difficult to get to any conclusion without a proper spectroscopic analysis of LERGs. Unfortunately, the studies of outflows in LERGs are mostly single object based and there is almost no systematic study on the outflows in [\ion{O}{3}]~$\lambda\lambda 4959, 5007$. We therefore pursue a systematic spectroscopic analysis to investigate the outflows in ionized gas.

In this paper, we focus on estimating the physical properties and determining the potential driving mechanism of the warm ($\sim 10^{4}~\text{K}$), ionized outflows in LERGs and to see the effect of these outflows on their host galaxies. We present an SDSS spectroscopic analysis of a sample of 802 LERGs with $0.01 < z < 0.3$ drawn from the parent sample of \citet{best2012}. Consequently, it implies that we could put our spectroscopic observations into the context of the entire nearby LERG population. 
In Section~\ref{sec:obs}, we first briefly describe our sample-selection criteria and the sample itself. We detail our method for detecting outflows in Section~\ref{sec:analysis_and_results}, while in Section~\ref{sec:discussion} we attempt to explain the key driving force behind the launching of the outflows, and what role these outflows play in terms of AGN feedback. We summarize our conclusions in Section~\ref{sec:summary_and_conclusions}. Throughout this paper we adopt the standard $\Lambda$CDM cosmology with $H_0 = 70~\text{km s}^{-1}~\text{Mpc}^{-1}$, $\Omega_{\text{m}} = 0.3$, and $\Omega_{\Lambda} = 0.7$.

\section{Data}\label{sec:obs}

We constructed this sample primarily from the spectroscopic sample from the 7th data release of the Sloan Digital Sky Survey (SDSS DR7) \citep[]{abazajian2009} and then combining SDSS data with the data from NVSS survey \citep[][]{condon1998}- The NRAO (National Radio Astronomy Observatory) VLA (Very Large Array) Sky Survey and the FIRST survey - The Faint Images of the Radio Sky at Twenty Centimeters  survey. The parent sample of SDSS DR7 consists of 927,552 galaxies in the form of value-added catalogues that include all the host galaxy and emission line properties properties, emission line and was constructed by the research groups from in Max-Planck Institute for Astrophysics (MPA) and John Hopkins University (JHU) \citep[e.g.,][]{kauffmann2003a,kauffmann2003b,brinchmann2004,tremonti2004}\footnote{\url{https://wwwmpa.mpa-garching.mpg.de/SDSS/DR7/}}. Following the prescription of \citet{best2005}, we cross-matched these galaxies with the radio sources from the NVSS and FIRST surveys. We further used the method described by \citet{donoso2009} in order to detect radio sources, not having FIRST counterparts. 

We initially retrieve radio information from the data catalogue of \citet{best2012}. This data catalogue consists of the characterization of the 18,286 radio sources, among which 9863 radio sources are LERGs. Explicitly, for this work we limit ourselves to LERGs with signal-to-noise ratio ($S/N$) $> 3$ for each of the following emission lines- H$\beta$, [\ion{O}{3}]~$\lambda 5007$, H$\alpha$, [\ion{N}{2}]~$\lambda 6584$, and [\ion{S}{2}]~$\lambda 6716, 6731$ for confirmed detection. We additionally select only LERGs with $z < 0.3$ to ensure that the H$\alpha$ line does not go out of the SDSS wavelength coverage.

\citet{buttiglione2010} defined a parameter, called `excitation index':
\begin{align*}
    \text{EI} = & \log{ \left( \frac{\text{[\ion{O}{3}]}}{\text{H}\beta} \right)} \\
                & -\frac{1}{3} \left[ \log{ \left( \frac{\text{[\ion{N}{2}]}}{\text{H}\alpha} \right)} +
                                      \log{ \left( \frac{\text{[\ion{S}{2}]}}{\text{H}\alpha} \right)} +
                                      \log{ \left( \frac{\text{[\ion{O}{1}]}}{\text{H}\alpha} \right)} \right].
\end{align*}
They demonstrated this parameter to be bimodal and mentioned that the approximate separation between LERGs and HERGs is around $\text{EI} \sim 0.95$. The lack of [\ion{O}{1}] information restricts us estimating $\text{EI}$ to select the LERGs. However, we adopt the classification of LERGs from the diagnostic diagrams of \citet{buttiglione2010} based on emission lines that are available:
\begin{align*}
    \log \left( \text{[\ion{O}{3}]}/\text{H}\beta \right) - \log \left( \text{[\ion{N}{2}]}/\text{H}\alpha \right) \lesssim 0.7, \\
    \log \left( \text{[\ion{O}{3}]}/\text{H}\beta \right) - \log \left( \text{[\ion{S}{2}]}/\text{H}\alpha \right) \lesssim 0.9.
\end{align*}

In order to trace outflow signatures in highly ionized lines we require the emission lines to be strong enough for the detection beyond the noise level. We also aim for a single consistent approach unlike \citet{best2012} where they carried out a multiple approach, combining different methods to classify the radio AGN until a classification was achieved. Consequently, we confine our analysis to radio sources which could be selected as LERGs by the analysis of \citet{best2012} and \citet{buttiglione2010}. We therefore only choose the sources which are identified as LERGs by both \citet{best2012} and \citet{buttiglione2010}. We further put special emphasis on their accretion rates. LERGs have been shown to have weak accretion rates \citep{best2012}. As the accretion rate goes up, beyond $1\%$ of the Eddington accretion rate, the radiation will become more efficient to drive the feedback. Therefore, we selected the LERGs which have Eddington ratio $\lambda_{\text{Edd}} < 0.01$. This selection process gives us 802 LERGs.

We use the data from the AGN Line Profile and Kinematics Archive \citep[ALPAKA,][]{mullaney2013} catalogue in order to put our results in context with respect to the work by previous spectral studies connecting outflows and radio sources. The ALPAKA catalogue consists of spectroscopic measurements of 24,264 type~1 and type~2 AGN \citep[Seyferts only -- LINERs were excluded;][]{mullaney2013}, including both strongly accreting ($\lambda_{\text{Edd}} > 0.01$) and weakly accreting ($\lambda_{\text{Edd}} < 0.01$) AGN. We additionally use imaging from FIRST to determine the radio morphology of our LERG sample.

    \begin{figure*}
        \centering
        \includegraphics[width=\textwidth]{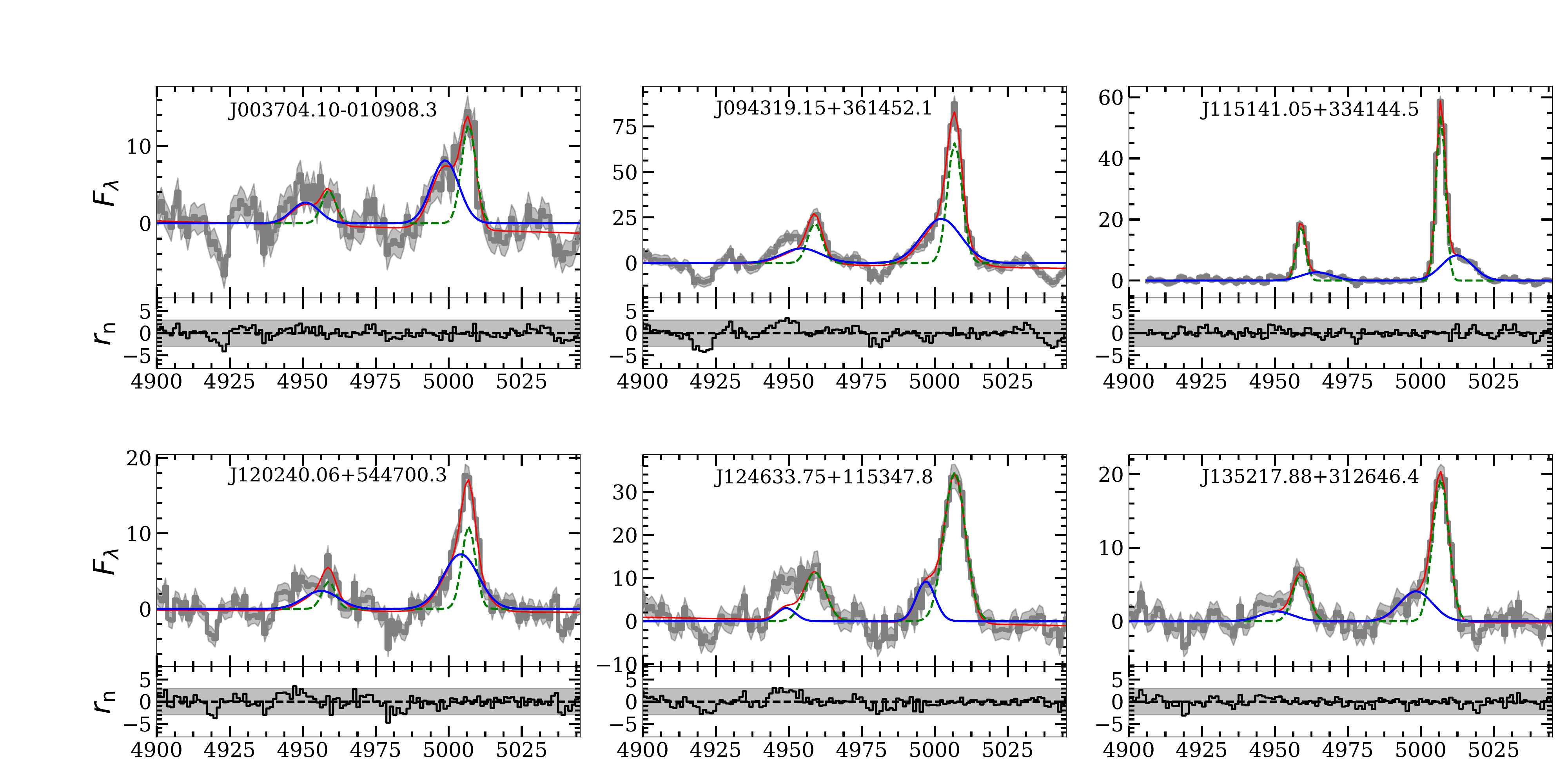}
        \includegraphics[width=\textwidth]{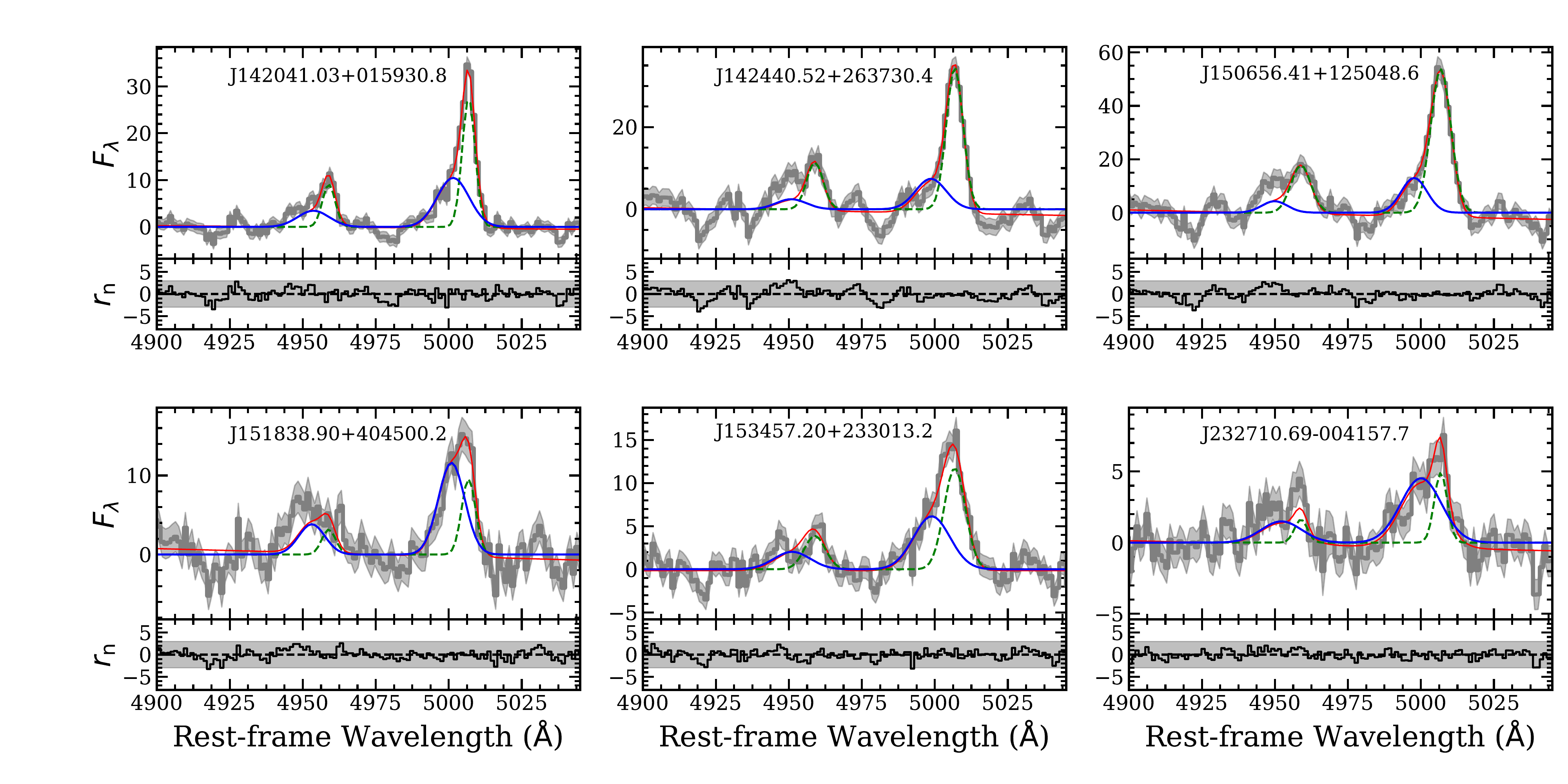}
        \caption{Multi-component modelling for the 12 weakly accreting LERGs studied here. Fits are arranged such that right ascension increases from left to right, top to bottom. \textit{Upper panels}: The SDSS spectrum and the best-fit spectral model, along with the breakdown of the fitted Gaussian components. The green Gaussian represents the narrow first Gaussian component (fit to the narrow core components of [\ion{O}{3}]) while the blue Gaussian represents the broad second Gaussian component (the best fit to the broad wing components of [\ion{O}{3}]). The total model of core and wing is denoted by the red line, while the data is shown in dark gray with $1 \sigma$ errors in light gray. The y-axis ($F_{\lambda}$) indicates the respective flux-density corresponding to the wavelengths given in units of $10^{-17}~\text{erg s}^{-1}~\text{cm}^{-2}~\text{\AA}^{-1}$. \textit{Lower panels}: Residual spectrum normalized by the error spectrum. The y-axis ($r_{\text{n}}$) stands for these normalized residuals. The dark grey shaded area highlights the $\pm 3 \sigma$ limit and the dashed line provides the reference to zero.}
        \label{fig:central_spectra_weak_accreters}
    \end{figure*}

    \begin{figure*}
        \centering
        \includegraphics[width=\textwidth]{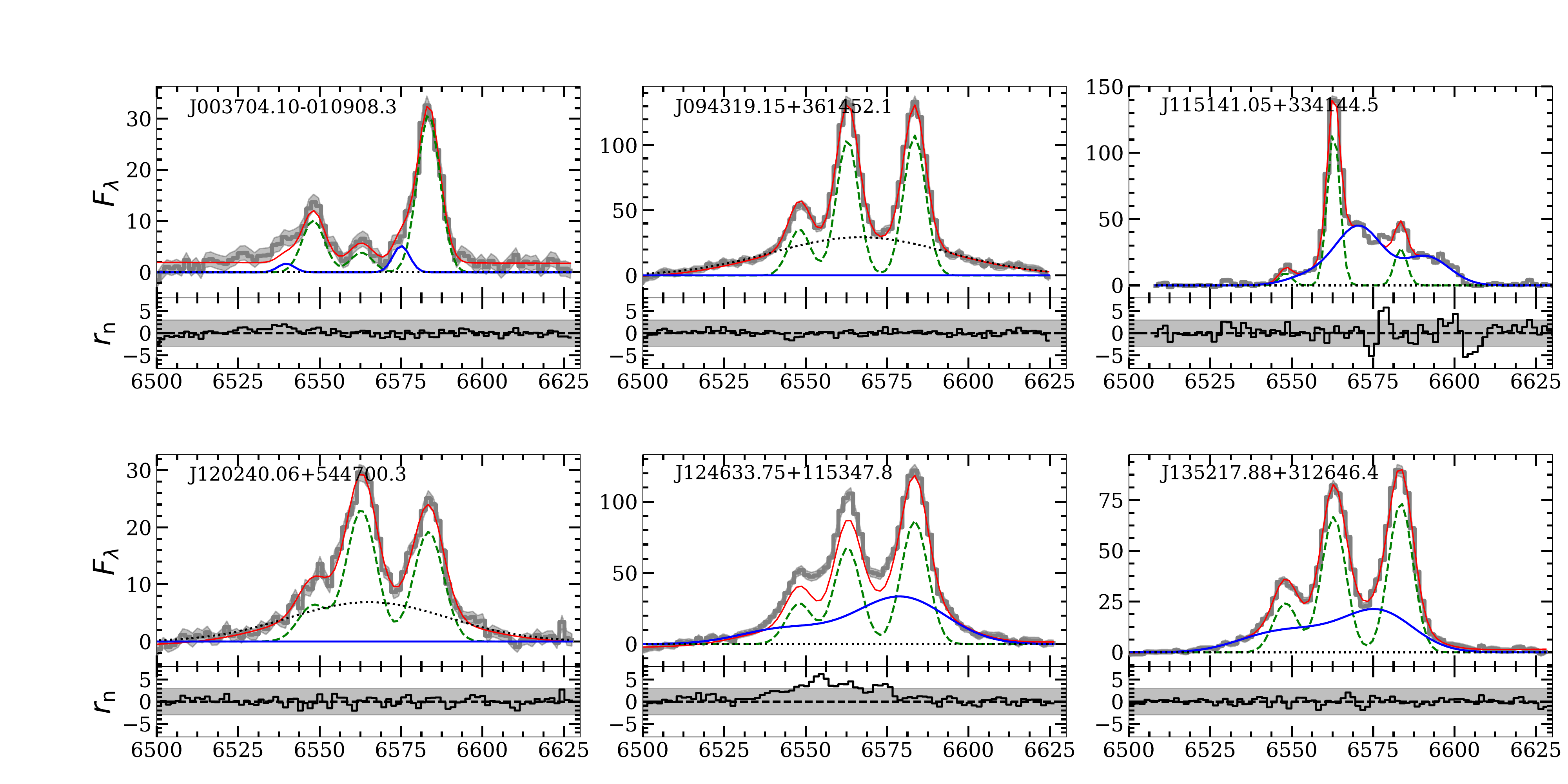}
        \includegraphics[width=\textwidth]{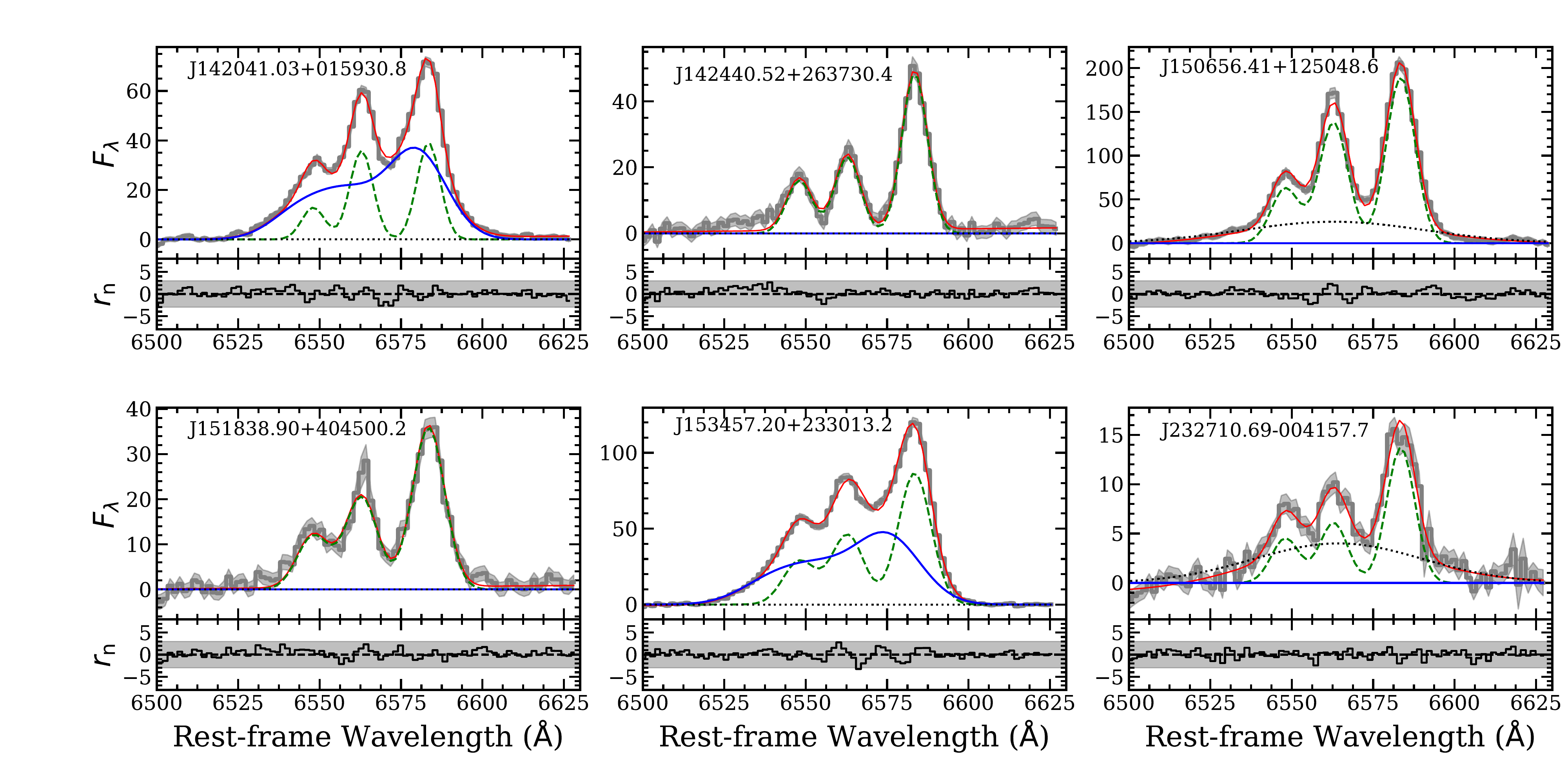}
        \caption{Multi-component modelling of the H$\alpha$+[\ion{N}{2}] complex for the 12 weakly accreting LERGs, arranged as in Figure~\ref{fig:central_spectra_weak_accreters}. Here the green Gaussians represent the fit to the H$\alpha$+[\ion{N}{2}] core, the blue Gaussians represent the fit to the wing component, while the dotted black curves denote the fit to the broad H$\alpha$ component. The red and grey curves represent the total model and data (along with shaded $1 \sigma$ errors), while the bottom panels are as in Figure~\ref{fig:central_spectra_weak_accreters}.}
        \label{fig:central_spectra_weak_accreters_Ha}
    \end{figure*}

    \begin{figure*}
        \centering
        \includegraphics[width=\textwidth]{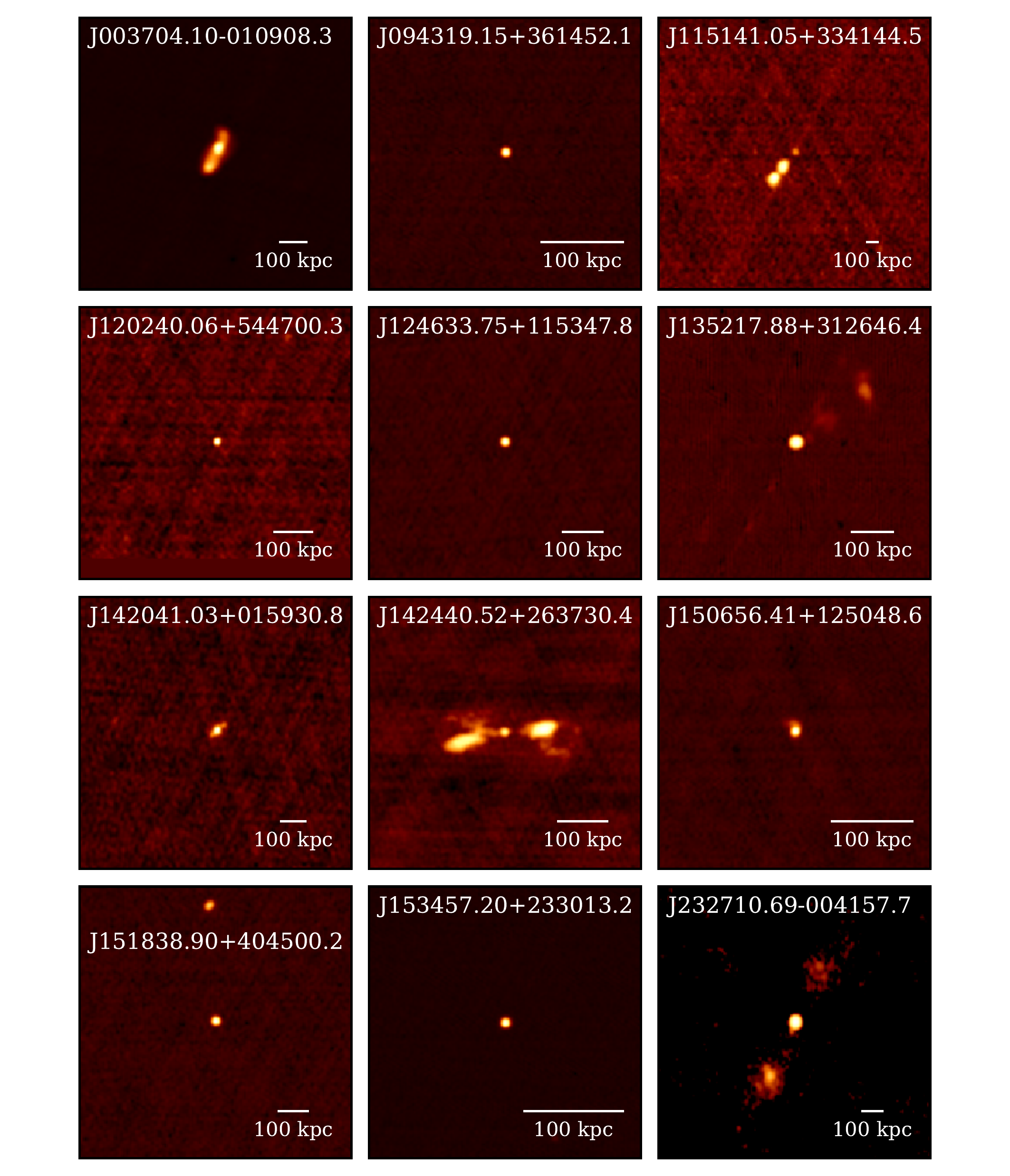}
        \caption{Total intensity radio images, as taken from the Faint Images of the Radio Sky at Twenty Centimeters \citep{becker1995} survey for the 12 weakly accreting LERGs with convincing blue wings in [\ion{O}{3}]. Images are arranged as in Figure~\ref{fig:central_spectra_weak_accreters}. A scale bar is present in each frame showing $100~\text{kpc}$ projected at the redshift of each radio source.}
        \label{fig:FIRST_images_weak_accreters}
    \end{figure*}

\section{Analysis and Results}\label{sec:analysis_and_results}

\subsection{[\ion{O}{3}] profile modelling}\label{subsec:[OIII]_model}

    \begin{deluxetable*}{cCCCCCCCc}
        \tablecaption{Properties of the weakly-accreting LERGs showing outflows.\label{tab:weak_accreters}}
        \tablehead{\colhead{Object} \vspace{-0.15cm} & \colhead{$z$} & \colhead{$\log(L_{\text{[O III]}}$)\tablenotemark{a}} & \colhead{$\log(L_{\text{wing}}$)\tablenotemark{b}} & \colhead{$\log(L_{1.4~\text{GHz}}$)} & \colhead{$\log(\lambda_{\text{Edd}}$)} & \colhead{$\Delta v$} & \colhead{$\sigma_{\text{wing}}$\tablenotemark{c}} & \colhead{Morphology} \\
                   & & \colhead{[$\text{erg s}^{-1}$]} & \colhead{[$\text{erg s}^{-1}$]} & \colhead{[$\text{W Hz}^{-1}$]} & & \colhead{[$\text{km s}^{-1}$]} & \colhead{[$\text{km s}^{-1}$]} &
                   }
        \startdata
        J003704.10-010908.3 & 0.074 & 40.74 & 40.50 & 25.70 & -2.19 & -404 $\pm$ 195 & 413 $\pm$ 97 & FR~I \\
        J094319.15+361452.1 & 0.022 & 40.34 & 40.05 & 22.93 & -2.12 & -160 $\pm$ 78 & 356 $\pm$ 47 & Compact \\
        J115141.05+334144.5 & 0.214 & 41.96 & 41.56 & 24.82 & -2.15 & -482 $\pm$ 81 & 276 $\pm$ 95 & FR~I \\
        J120240.06+544700.3 & 0.049 & 40.37 & 40.19 & 22.49 & -2.11 & -509 $\pm$ 86 & 329 $\pm$ 37 & Compact \\
        J124633.75+115347.8 & 0.047 & 40.69 & 39.97 & 23.51 & -2.45 & -385 $\pm$ 309 & 290 $\pm$ 185 & Compact \\
        J135217.88+312646.4 & 0.045 & 40.34 & 39.83 & 25.35 & -2.39 & -490 $\pm$ 105 & 306 $\pm$ 29 & FR~II \\
        J142041.03+015930.8 & 0.078 & 41.02 & 40.73 & 23.25 & -2.16 & -595 $\pm$ 201 & 185 $\pm$ 100 & FR~I \\
        J142440.52+263730.4 & 0.037 & 40.42 & 39.88 & 24.36 & -2.16 & -544 $\pm$ 91 & 268 $\pm$ 196 & FR~II \\
        J150656.41+125048.6 & 0.022 & 40.22 & 39.60 & 23.00 & -2.61 & -547 $\pm$ 197 & 258 $\pm$ 79 & Compact \\
        J151838.90+404500.2 & 0.065 & 40.67 & 40.52 & 23.65 & -2.07 & 330 $\pm$ 79 & 310 $\pm$ 39 & Compact \\
        J153457.20+233013.2 & 0.018 & 39.57 & 39.25 & 23.40 & -2.88 & -275 $\pm$ 87 & 399 $\pm$ 44 & Compact \\
        J232710.69-004157.7 & 0.099 & 40.81 & 40.69 & 24.75 & -2.07 & -663 $\pm$ 84 & 261 $\pm$ 74 & FR~II \\
        \enddata
        \tablenotetext{a}{Logarithm of the [\ion{O}{3}]~$\lambda 5007$ luminosity, which includes the luminosity from each of the two fitted Gaussian components.}
        \tablenotetext{b}{Logarithm of the luminosity calculated from the broad Gaussian component fitted to [\ion{O}{3}]~$\lambda 5007$. We assume this indicates the luminosity of the outflowing [\ion{O}{3}]~$\lambda 5007$ gas.}
        \tablenotetext{c}{Velocity dispersion of the broad Gaussian component fitted to [\ion{O}{3}]~$\lambda 5007$. We assume this indicates the line-width of the outflowing [\ion{O}{3}]~$\lambda 5007$ gas.}
    \end{deluxetable*}

We use the stellar continuum-subtracted spectra provided by SDSS throughout the analysis. As we look for the signs of outflows in [\ion{O}{3}] line emitting gas, we mainly focus on modelling the [\ion{O}{3}]~$\lambda 5007$ line-shape. In order to get a better constraint on the fit, we extended our spectral modelling to [\ion{O}{3}]~$\lambda 4959$. We therefore consider the rest-frame wavelength region from $4900~\text{\AA}$ to $5050~\text{\AA}$ which covers the [\ion{O}{3}]~$\lambda\lambda 4959, 5007$ doublet. This large wavelength window enables us to account for all the spectral lines covering narrow emission line region (NLR) components, especially any asymmetry in the line-shape (if present) which could potentially affect our measurement of [\ion{O}{3}] kinematic parameters. 

In order to implement our spectral model, we assume that the velocity centers and the line widths of the narrow [\ion{O}{3}] $\lambda\lambda4959,5007$ are exactly the same, because they likely originate from the same physical region with similar kinematics. We further put two constraints on the line fluxes in addition to our previous assumptions of emission line kinematics. The first constraint states that the [\ion{O}{3}]~$\lambda\lambda 4959, 5007$ complex has a line flux ratio of $1/3$ \citep{storey2000} as expected from subatomic physics. Secondly, we approximated the continuum with a linear model as the wavelength range of interest was a very small part of the wavelength range covered by the spectrograph. In order to use only high $S/N$ lines and clarify potential asymmetries we further calculated the average $S/N$ within the wavelength range from $4997~\text{\AA}$ to $5017~\text{\AA}$ which contains the emission line [\ion{O}{3}]~$\lambda 5007$. If $S/N > 3$ we proceeded with modelling the data.

We model the [\ion{O}{3}]~$\lambda\lambda 4959, 5007$ doublet using a non-linear Levenberg-Marquardt algorithm with a single or double Gaussian function accounting for each of the emission lines. We only adopt a double Gaussian profile if the skewness within the rest-frame wavelength range from $4980-5025~\text{\AA}$ is outside the range $[-0.5, 0.5]$. This shows a prominent asymmetry in the [\ion{O}{3}] line profile. We assume that the line-center of the narrowest [\ion{O}{3}]~$\lambda 4959, 5007$ component (core) lies exactly at the host galaxy's systematic redshift.
\citet{mullaney2013} found that that the redshift provided by the SDSS data base, agrees very well with the redshift derived from the [\ion{O}{3}] core.
We define the velocity shift, $\Delta v$, as $\Delta v$ = $v_{2} - v_{1}$, where $v_{1}$ and $v_{2}$ are the line centers of the first (core) and second Gaussian (wing) components, respectively. A velocity shift $\Delta v > 0$ indicates the presence of a blue-wing, while a red-wing has $\Delta v < 0$. Furthermore, we create 100 mock spectra by fluctuating the spectroscopic data with their respective uncertainties and repeat the modeling process, and then estimate the uncertainty in individual model parameters by taking the standard deviation of the distribution. 
In order to avoid erroneously fitting the noise as the second Gaussian component, we adopt the following criterion:

\begin{itemize}
    \item If the amplitude of a Gaussian component is $A$ with an associated uncertainty $N$, then both Gaussian components must have $A/N > 3$
    
    \item The fractional error in velocity dispersion in each of the Gaussian components must be $< 1$.
    
\end{itemize}

We describe the second Gaussian component to indicate an outflow only if the fractional error in $\Delta v< 1$.

In summary, we only describe the visible asymmetry in the [\ion{O}{3}]~$\lambda\lambda 4959, 5007$ doublet as being indicative of an outflow when both fitted Gaussians are significant and they are offset from each other beyond the uncertainty level. This selection process results in 14 sources with confirmed outflows in [\ion{O}{3}], with 13 sources showing visible blue-wings and one source showing a red-wing.

\citet{crenshaw2010} explained whether there are instances where the outflow geometry and extinction cumulatively can result in a red wing or red wards asymmetry in emission line (redshifted emission line). They showed a test case where the position angle (PA) and inclination of the galactic disk are such that the redshifted cone in the north is not totally occulted by the galactic disk but the blueshifted cone located in the southern side, is fully occulted by the galactic disk. They further ran a simulation for each of the possible combinations of bi-cone inclination ($i$), inclination of the disk, and the difference in the position angle (in $1\arcdeg$ intervals) which were further weighted by the probability of observing a bi-cone at a given $i$ ($\propto \sin{i}$); and assumed a random distribution of these parameters. \citet{crenshaw2010} found that the percentage of the total population, showing more extinction of the red wing (which is the redshifted portion of the bi-cone) than the blue wing (which is the blueshifted portion of the bi-cone) is $17.2\%$, $16.7\%$, $15.6\%$, and $13.7\%$ when the bi-cone's half opening angles are $30\arcdeg$, $40\arcdeg$, $50\arcdeg$, and $60\arcdeg$, respectively \citep{crenshaw2010}. They concluded that their model could explain the relatively small occurrences of redshifted emission lines.

In some spectra, we notice a `dip' in the wavelength rest-frame wavelength region from $4970-4990~\text{\AA}$ and in some cases the `dip' is comparable to the neighboring emission line [\ion{O}{3}]~$\lambda 5007$. The presence of this `dip' essentially indicates fluctuations in noise in the spectra, and therefore any results from this sub-sample will be unreliable. To filter out the results from these spectra, we define two regions: region~1, where the rest-frame wavelength is between $4970-4990~\text{\AA}$, and region~2, where the rest-frame wavelength is between $4995-5015~\text{\AA}$. Region~2 contains the emission line region. If the absolute value of the amplitude in region~2 was more than three times of that in region~1, then we considered it a reliable result. This process eliminates two additional sources.

Consequently, we find 12 LERGs with outflows, representing $\sim 1.5\%$ of the parent sample of 802 LERGs. Plots of the model fits to the nuclear spectra for these 12 LERGs can be seen in Figure~\ref{fig:central_spectra_weak_accreters}, with derived parameters tabulated in Table~\ref{tab:weak_accreters}.

\subsection{H$\alpha$ + [\ion{N}{2}] profile modelling}\label{subsec:Ha_model}

We model the H$\alpha$ + [\ion{N}{2}]~$\lambda\lambda 6548, 6583$ doublet using a non-linear Levenberg-Marquardt algorithm with a multi-Gaussian function accounting for each of the emission lines, as seen in Figure~\ref{fig:central_spectra_weak_accreters_Ha}. The H$\alpha$ + [\ion{N}{2}] complex shows a great diversity among the 12 sources. In some sources the line-profile does not show any asymmetry, whereas in some cases there is a prominent asymmetry. Further, in some cases we notice a visible broadening in the spectral shape of H$\alpha$ suggesting the possible existence of a broad line region (BLR).

We primarily adopt a double Gaussian model to check for outflows in H$\alpha$ + [\ion{N}{2}]~$\lambda\lambda 6548, 6583$ as seen in their [\ion{O}{3}] line shape, where each of the Gaussian triplets represent core and wing components. We again assume that the velocity centers and the line widths of the [\ion{N}{2}]~$\lambda\lambda 6548, 6583$ and H$\alpha$ are exactly the same for each of the core and wings. Additionally, we assume that the [\ion{N}{2}]~$\lambda\lambda 6548, 6583$ complex has a line flux ratio of $1/3$ based on subatomic physics, and again use a linear model to account for the continuum. We confirm the necessity of the second Gaussian if both of the Gaussian components have $A/N > 3$ and the fractional error in velocity dispersion is less than unity. In {J003704.10$-$010908.3} and {J124633.75$+$115347.8}, the H$\alpha$ wing component has $A/N < 3$ but $A/N > 3$ in the [\ion{N}{2}]~$\lambda\lambda 6548, 6583$ wings. Therefore, we neglect the H$\alpha$ wing component in those two sources. In four objects, we see that the FWHM of the second Gaussian component of H$\alpha$ is $>2000~\text{km s}^{-1}$ with $A/N < 3$ for the second Gaussian component of [\ion{N}{2}]. For these cases, we refit the spectra with a single Gaussian model accounting for all the narrow cores of H$\alpha$ + [\ion{N}{2}]~$\lambda\lambda 6548, 6583$, and a broad Gaussian accounting for the H$\alpha$ BLR. We checked that each of the Gaussian components has $A/N > 3$ and the fractional error in velocity dispersion is less than unity. In {J142440.52$+$263730.4} and {J151838.90$+$404500.2}, we see that both the H$\alpha$ and [\ion{N}{2}]~$\lambda\lambda 6548, 6583$ wings have $A/N > 3$. Therefore, we do not use any further Gaussian component beyond the Gaussian triplet representing the H$\alpha$ + [\ion{N}{2}]~$\lambda\lambda 6548, 6583$ core.

\subsection{Star formation rates (SFR) for the LERGs with outflows}\label{subsec:SFR}

We estimate the SFR for those LERGs with outflows using the H$\alpha$ core luminosity as derived from the H$\alpha$ + [\ion{N}{2}] spectral modelling using the relation from \citet{kennicutt1998}:
\begin{equation}
    \frac{\text{SFR}}{M_{\odot}~\text{yr}^{-1}} = 7.9 \times 10^{-42} \frac{L_{\text{H}\alpha}}{\text{erg s}^{-1}}.
\end{equation}
Derived SFRs are listed in Table~\ref{tab:lerg_feedback}.

\subsection{Radio morphology}\label{subsec:radio_morph}

We additionally retrieved radio images from FIRST \citep{becker1995} for the 12 weakly accreting LERGs with ionized outflows seen in [\ion{O}{3}]. In Figure~\ref{fig:FIRST_images_weak_accreters} we present total intensity images showing the radio morphology of each object. We find that six of our 12 weakly accreting LERGs are Fanaroff-Riley \citep[FR,][]{fanaroff1974} class~0 objects (compact sources), three are FR~I, and three are FR~II with faint lobes. The source properties are compiled in Table~\ref{tab:weak_accreters}. 

\subsection{Mass outflow rates}\label{subsec:mass_outflow_rates}

The key parameters that are widely used to investigate the  role that AGN driven outflows play in their host galaxies' evolution are the mass, momentum and the energy carried by these galactic-scale outflows. Outflows are observed to be multi-phase phenomena \citep[e.g.,][]{shih2010,hardcastle2012,mahony2013,rupke2013} containing the hot outflows seen in X-ray \citep[e.g.,][]{tombessi2010,tombessi2011}, the fast-outflows seen in UV \citep[e.g.,][]{arav2015,arav2020}, the warm ionized outflows seen in optical \citep[e.g.,][]{liu2013a,liu2013b,harrison2014,woo2016}, neutral \citep[e.g.,][]{morganti2005a,morganti2005b,emonts2005} and molecular \citep[e.g.,][]{dasyra2014,cicone2014} which means that these outflows could entrain gas across all these multiple-phases. Therefore, it is necessary to acquire observations across all these wavelength ranges to understand the physical properties of these outflows. In this paper, we focus on the warm ionized phase of these outflows as observational data only covers the visible wavelength range. As found by \citet{rupke2013}, these warm, ionized phase of the outflows could account for a large fraction of the total mass, momentum and energy carried out by the multi-phase outflows. One challenge is to constrain the geometry of these outflows as pointed out by \citet{harrison2014,husemann2016}.

The NLR ionized gas clouds are often driven in outflows by the central engine \citep{hutchings1998,crenshaw2000a,crenshaw2000b,fischer2013} and usually tend to have bi-conical structures, where the apex of the bi-cone resides in the central AGN \citep{pogge1988,schmitt1994}. In recent years \citet{fischer2013} employed a bi-conical geometry on \textit{Hubble Space Telescope} long-slit spectroscopic data for the NLR gas clouds, with both of the cones being identical. They found that the bi-conical geometry provides the best explanation behind how [\ion{O}{3}] images appear to be show axisymmetric often, triangular NLRs for Seyfert~2s and whereas, for Seyfert~1s, the NLRs appear to be compact circular or elliptical NLRs. This picture is consistent with the unified model of AGN \citep{schmitt2003}. 

\citet{canodiaz2012} and \citet{cresci2015} suggested a bi-conical outflow geometry and prescribed a simplistic model where the bi-cone is uniformly filled with outflowing ionized gas cloud, in their work on high redshift ($z > 2$) quasars. \citet{canodiaz2012} reported of highly blueshifted regions southwards and eastwards to the nucleus of the luminous quasar {2QZJ002830.4-281706}, appearing to a form a 'bow-like morphology', indicative of the envelope of a prominent bi-conical outflow. \citet{canodiaz2012} and \citet{cresci2015} assumed a conical geometry with a given opening angle, uniformly dense, uniformly distributed ionized gas clouds with a constant outflow velocity.
\citet{canodiaz2012} mentioned that the outflow could be symmetric but the receding side of the outflow or the red wind could be obscured by the disk of the host galaxy.
Assuming ionized gas clouds with constant density, case~B recombination \citep[e.g.,][]{rupke2013} and an electron temperature, $T\sim 10^{4}~\text{K}$, one can easily estimate the mass outflow rates ($\dot{M}_{\text{ion}}$) and kinetic energy outflow rates ($\dot{E}_{\text{kin}}$). The advantage of this assumption in that $\dot{M}_{\text{ion}}$ and $\dot{E}_{\text{kin}}$ become independent of the filling factor of the ionized gas clouds within the bi-cone and the opening angle. \citet{husemann2016} derived the following relations:
\begin{equation}\label{eq:M_out}
    \begin{split}
    \frac{\dot{M}_{\text{ion}}(D)}{3~M_{\sun}~\text{yr}^{-1}} = & \left( \frac{L_{H\beta}}{10^{41}~\text{erg s}^{-1}} \right) \left( \frac{n_{\text{e}}}{100~\text{cm}^{-3}} \right)^{-1} \times \\
    & \left( \frac{v_{\text{out}}}{100~\text{km s}^{-1}} \right) \left( \frac{D}{\text{kpc}} \right)^{-1};
    \end{split}
\end{equation}

\begin{equation}\label{eq:E_kin}
     \frac{\dot{E}_{\text{kin}}(D)}{10^{40}~\text{erg s}^{-1}} = \left(  \frac{\dot{M}_{\text{ion}}(D)}{3~M_{\sun}~\text{yr}^{-1}} \right) \left( \frac{v_{\text{out}}}{100~\text{km s}^{-1}} \right)^{2},
\end{equation}
where $v_{\text{out}}$ is the outflow velocity. As H$\beta$ has much lower $S/N$ than [\ion{O}{3}] for these LERGs, it is difficult to obtain the luminosity of the H$\beta$ wing component. In fact, even the H$\beta$ core component is difficult to determine. We therefore assume that the [\ion{O}{3}]/H$\beta$ ratio for the wing components would be similar to their flux ratios determined by the SDSS pipeline, and obtain the H$\beta$ wing luminosity by simply multiplying the [\ion{O}{3}] luminosity by the [\ion{O}{3}]/H$\beta$ flux ratio from the MPA/JHU catalogue.

One important quantity for estimating the effect of feedback is the outflow size, $R_{\text{out}}$. Unfortunately, one-dimensional spectroscopic data does not allow us to directly estimate the possible sizes of the outflow. \citet{kang2018} kinematically measured outflow sizes by estimating the distance from the nucleus where the [\ion{O}{3}] velocity dispersion becomes equal to the host-galaxy's stellar velocity dispersion using Gemini Multi-Object Spectrographic data. They determined a relation between outflow size and [\ion{O}{3}] luminosity which we employ to establish an upper limit on the outflow sizes from these LERGs \citep{kang2018}:
\begin{equation}
    \begin{split}
        \log{ \left( \frac{R_{\text{int}}}{\text{pc}} \right) } = & (0.250 \pm 0.018) \log{ \left( \frac{L_{\text{[O III]}}}{10^{42}~\text{erg s}^{-1}} \right) } \\
        & + (3.746 \pm 0.028).
    \end{split}
\end{equation}

Another necessary quantity for measuring ionized gas masses, and hence mass outflow rates, is the electron density, $n_{\text{e}}$. As the wing component of [\ion{S}{2}] was not detected during our analysis, we use the electron density obtained by \citet{kakkad2018}, where $n_{\text{e}}$ drops exponentially with increasing distance from the nucleus. However, this relation requires a maximum possible $n_{\text{e}}$ representing the $n_{\text{e}}$ near the AGN. \citet{kakkad2018} found that the electron-density for the outflowing gas is $\sim 50 - 2000~\text{cm}^{-3}$. We therefore use $2000~\text{cm}^{-3}$ as the maximum allowable electron density, and estimate the electron density at $R_{\text{out}}$. We estimate the mass and kinetic energy outflow rates for the LERGs with outflows using Equations~\ref{eq:M_out} and \ref{eq:E_kin} which we list in Table~\ref{tab:lerg_feedback}.

    \begin{deluxetable*}{cCCCCCCCc}
        \tablecaption{Feedback parameters of the LERGs with outflows.\label{tab:lerg_feedback}}
        \tablehead{\colhead{Object} \vspace{-0.15cm} & \colhead{$\log(L_{\text{H}\alpha_{\text{core}}}$)\tablenotemark{a}} & \colhead{$\text{SFR}$} &
        \colhead{$\log(\text{sSFR})$} &
        \colhead{$\log(R_{\text{out}}$)} & \colhead{$\log(\dot{E}_{\text{out}}$)\tablenotemark{b}} & \colhead{$\dot{M}_{\text{out}}$\tablenotemark{c}} & \colhead{$\eta$\tablenotemark{d}} \\
                   & \colhead{[$\text{erg s}^{-1}$]} & \colhead{[$M_{\sun}~\text{yr}^{-1}$]} & \colhead{[$\text{yr}^{-1}$]} & \colhead{[$\text{kpc}$]} & \colhead{[$\text{erg s}^{-1}$]} & \colhead{[$M_{\sun}~\text{yr}^{-1}$]} & &
                   }
        \startdata
        J003704.10$-$010908.3 & 41.29 & 1.54 & -10.63 & 1.19 & 42.73 & 97.77 & 63.6\\
        J094319.15$+$361452.1 & 41.63 & 3.39 & -10.67 & 0.92 & 42.27 & 45.62 & 13.5\\
        J115141.05$+$334144.5 & 43.47 & 232.19 & -11.80 & 2.61 & 42.05 & 131.62 & 0.6\\
        J120240.06$+$544700.3 & 41.83 & 5.29 & -12.10 & 0.94 & 42.65 & 59.74 & 11.3\\
        J124633.75$+$115347.8 & 42.20 & 12.56 & -11.41 & 1.15 & 42.44 & 35.41 & 2.8\\
        J135217.88$+$312646.4 & 42.12 & 10.45 & -11.43 & 0.92 & 42.50 & 36.82 & 3.5\\
        J142041.03$+$015930.8 & 42.33 & 16.79 & -12.22  & 1.43 & 42.77 & 175.58 & 10.5\\
        J142440.52$+$263730.4 & 41.48 & 2.53 & -12.13 & 0.97 & 42.71 & 63.94 & 25.3\\
        J150656.41$+$125048.6 & 41.86 & 6.08 & -9.30 & 0.85 & 42.68 & 40.81 & 6.7\\
        J151838.90$+$404500.2 & 41.69 & 9.22 & -10.90 & 1.14 & 42.90 & 80.27 & 8.7\\
        J153457.20$+$233013.2 & 41.28 & 1.51 & -12.11 & 0.56 & 41.42 & 7.22 & 4.8\\
        J232710.69$-$004157.7 & 41.97 & 5.75 & -10.24 & 1.24 & 42.58 & 151.11 & 26.3\\
        \enddata
        \tablenotetext{a}{Logarithm of the narrow H$\alpha$ core luminosity, from the multi-Gaussian modeling of H$\alpha$ + [\ion{N}{2}] complex (see Section~\ref{subsec:Ha_model}).}
        \tablenotetext{b}{Logarithm of the kinetic energy carried out by the outflow per unit time (see Section~\ref{subsec:mass_outflow_rates}). We assume a bi-conical outflow geometry.}
        \tablenotetext{c}{Mass carried out by the outflow per unit time (see Section~\ref{subsec:mass_outflow_rates}).}
        \tablenotetext{d}{Mass loading factor as obtained from the ratio of mass outflow rate to star formation rate (SFR; ie. $\eta = \dot{M}_{\text{out}} / \text{SFR}$).}
    \end{deluxetable*}

\section{Discussion}\label{sec:discussion}

In this section, we put our results in the context of previous studies to investigate the possible driving force behind the outflows.

\subsection{Do compact radio sources preferentially drive outflows?}\label{subsec:compact_sources}

\citet{heckman1984} analyzed 34 moderate resolution ($\sim 70-150~\text{km s}^{-1}$) quasar spectra of the prominent [\ion{O}{3}]~$\lambda 5007$ emission line using data from three telescopes: the image-dissector scanner coupled to the Cassegrain spectrograph on the European Southern Observatory $3.6~\text{m}$ telescope, the High Gain Video Spectrometer coupled to the RC spectrograph on the Kitt Peak National Observatory $4~\text{m}$ Mayall telescope, and the intensified photon-counting Reticon spectrograph mounted on the Arizona-Smithsonian Astrophysical Observatory Multiple Mirror Telescope. They found that the [\ion{O}{3}]~$\lambda 5007$ line-width from the NLR are correlated with the luminosity of the portion of the steep-spectrum radio emission which occurs on a size scale $\sim 10^{2}-10^{3}~\text{pc}$. \citet{whittle1992} analyzed a sample of 140 Seyfert AGN with high radio luminosity ($L_{1.4~\text{GHz}} > 10^{22.5}~\text{W Hz}^{-1}$) and are linera radio sources. They found that these radio AGN have significantly larger FWHM ($>300~\text{km s}^{-1}$) in highly ionized [\ion{O}{3}] lines, suggesting that the jets could interact with the interstellar medium close to the nucleus. Long-slit observations of these radio AGN have revealed high-velocity components approximately coincident with the near-nuclear radio lobes, suggesting that the radio jet, that is responsible for powering the radio source does further accelerate the ionized gas, which cretaes a super-virial bipolar flow \citep{whittle1992}. Their studies also supported the idea of a compact radio-core which could possibly disturb the [\ion{O}{3}] to create such broad line-widths. Several spectral studies of relatively small samples ($<20$ sources) reported that, in case of radio AGN with compact radio jets reported, there are visible broadening in the [\ion{O}{3}]~$\lambda 5007$ emission line, co-spatial with these compact/small-scale radio jets \citep[e.g.,][]{gelderman1994,tadhunter2003,holt2006,holt2008}. They suggested that the mechanical energy from the radio jets could disturb the [\ion{O}{3}]~$\lambda 5007$ emitting gas clouds in these cases. An additional spectroscopic study by \citet{mullaney2013} examined 24,264 optically selected AGN with $z < 0.4$. Similar to the authors above, \citet{mullaney2013} reported a connection between the outflows and compact radio cores \citep[for a recent review, see][]{odea2021}.

\citet{mullaney2013} used the flux-weighted average FWHM ($FWHM_{\text{avg}}$) of the [\ion{O}{3}]~$\lambda 5007$ line in which the contributions from both fitted Gaussians could contribute rather than just the broad Gaussian component. $FWHM_{\text{avg}}$ is very useful in case there is any significant broadening ($\text{FWHM} > 500~\text{km s}^{-1}$) in the narrower Gaussian component. It is defined as:
\begin{equation}
    FWHM_{\text{avg}} = \sqrt{(FWHM_{\text{c}} f_{\text{c}})^2 + (FWHM_{\text{w}} f_{\text{w}})^2},
\end{equation}
where $f_{\text{c}}$ and $f_{\text{w}}$ are the fractional fluxes contained within the fitted Gaussian components (core and wing), and $FWHM_{\text{c}}$ and $FWHM_{\text{w}}$ are FWHMs of the core and wing components respectively. They stated that, focusing on the the flux-weighted average FWHM ($FWHM_{\text{avg}}$) would allow one to avoid arbitrary definitions such as the threshold above which the contribution from a broad component to the overall flux of the [\ion{O}{3}]~$\lambda 5007$ could be considered significant or the threshold beyond which a fitted Gaussian component could be called 'broad'. \citet{mullaney2013} mentioned that using $FWHM_{\text{avg}}$ would ensure all sources, if their [\ion{O}{3}]~$\lambda 5007$ line-profiles are modelled with either one or two Gaussians, could be compared on an equal footing (i.e. which includes, for example, AGN where the [\ion{O}{3}]~$\lambda 5007$ appears to be 'broad' but could be modelled with a single Gaussian component).

\begin{figure}[t]
        \centering
        \includegraphics[width=\columnwidth]{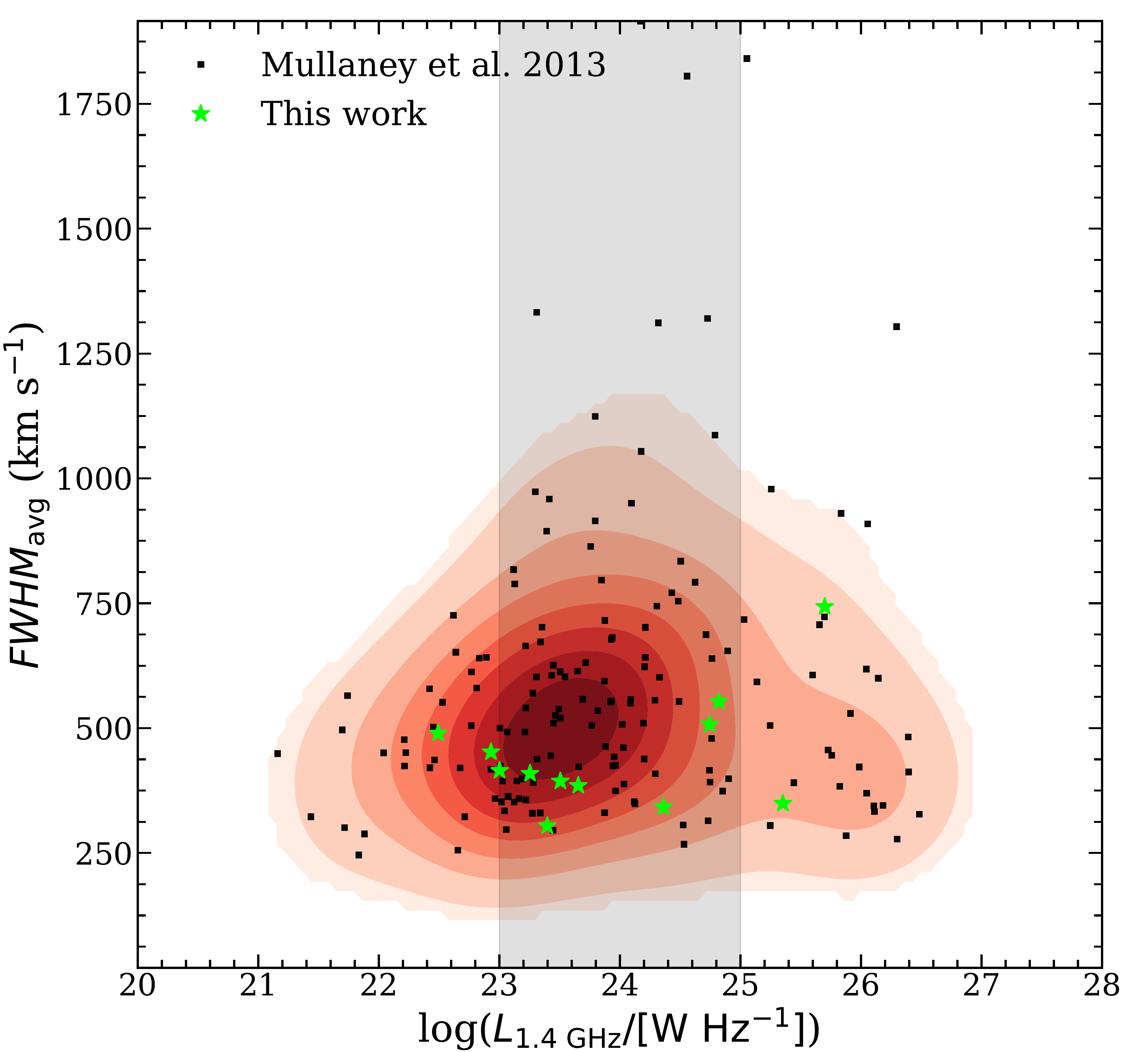}
        \caption{$FWHM_{\text{avg}}$ as a function of $1.4~\text{GHz}$ radio-luminosity ($L_{1.4~\text{GHz}}$) for our outflowing LERGs, and the AGN with $\lambda_{\text{Edd}} < 0.01$ from the work by \citet{mullaney2013}. Contours of the sample from \citet{mullaney2013} in the $FWHM_{\text{avg}}-L_{1.4~\text{GHz}}$ plane are overlaid to show any noticeable increase in $FWHM_{\text{avg}}$ within the radio-luminosity range $10^{23} < L_{1.4~\text{GHz}} < 10^{25}~\text{W Hz}^{-1}$. The gray colored shaded region indicates the radio-luminosity range $10^{23} < L_{1.4~\text{GHz}} < 10^{25}~\text{W Hz}^{-1}$.}
        \label{fig:FWHMavg_vs_Lrad}
    \end{figure}

In Figure~\ref{fig:FWHMavg_vs_Lrad}, we plot the flux-weighted average FWHM against the radio luminosity for our outflowing LERGs. We further over-plot the weakly accreting AGN ($\lambda_{\text{Edd}} < 0.01$) from \citet{mullaney2013}. We do not see that the $FWHM_{\text{avg}}$ of our LERGs increase significantly between the radio luminosity range $10^{23} < L_{1.4~\text{GHz}} < 10^{25}~\text{W Hz}^{-1}$, whereas the weak accretors in the sample of of \citet{mullaney2013} clearly show an increase in $FWHM_{\text{avg}}$ in that range. For the weakly accreting sources, the sample of \citet{mullaney2013} has a mean $FWHM_{\text{avg}} \sim 600~\text{km s}^{-1}$, whereas our outflowing LERGs have $FWHM_{\text{avg}} \sim 400~\text{km s}^{-1}$.

\citet{mullaney2013} reported that the [\ion{O}{3}]~$\lambda 5007$ line-width peaks between the radio luminosity $10^{23} < L_{1.4~\text{GHz}} < 10^{25}~\text{W Hz}^{-1}$. Those AGN with moderate-$L_{1.4~\text{GHz}}$ represent $\sim3\%$ of the optically selected AGN population at $z < 0.123$\footnote{NVSS is complete to $L_{1.4~\text{GHz}} > 10^{23}~\text{W Hz}^{-1}$ at $z < 0.123$.} in the sample of \citet{mullaney2013}. Despite constituting only $3\%$ of the optically selected AGN population, these moderate-$L_{1.4~\text{GHz}}$ AGN are $\sim 10$ times more likely to appear than the more radio luminous counterparts which are usually focused in the studies of AGN driven outflows \citep[e.g.,][]{holt2003,holt2008,tadhunter2003,nesvadba2006,nesvadba2008,canodiaz2012,harrison2012}.

\citet{mullaney2013} used the NVSS-detected AGN in their sample with $3 \times10^{23} < L_{1.4~\text{GHz}} < 3 \times 10^{24}~\text{W Hz}^{-1}$, covering the region around the $FWHM_{\text{avg}}$ peak. They reported that all 71 AGN in their sample with single FIRST matches and $FWHM_{\text{avg}} > 1500~\text{km s}^{-1}$ are unresolved (deconvolved extents $< 2~\text{arcsec}$), and the more extended radio sources do not have larger $FWHM_{\text{avg}}$ in [\ion{O}{3}]~$\lambda 5007$ (broader line). For the sources with multiple FIRST matches, $>80\%$ of the AGN in their subsample have at least one radio component closer than $1~\text{arcsec}$ to the central engine, which suggests the existence of a radio core. Additionally, they found that none of the AGN in their radio-selected subsample have multiple NVSS matches. \citet{mullaney2013} concluded that they are unlikely to be FR~I and FR~II sources.

We further investigate the fraction of compact radio sources in our sample of 802 LERGs. As per the work of \citet{best2005}, we define the radio sources which have single component in NVSS and a single FIRST match to be class~1; class~2 sources have single-component in NVSS, which is resolved into multiple components by FIRST; class~3 sources have single component in NVSS but they do not have a FIRST counterpart; and sources which are categorized into class~4 usually have multiple NVSS components. If we imagine the extended radio sources to be only contained in class~2, then the compact sources represent $\sim 98\%$ of the LERG population and $\sim 2\%$ of sources are extended radio sources. If we assume class~1 objects to be compact, and class~2, 3 and 4 to contain extended radio sources then the compact radio sources represent $87\%$ of the LERG population and the remaining $13\%$ of sources belong to the extended radio source population. We find for our outflowing LERGs that $50\%$ are compact radio sources (see Table~\ref{tab:weak_accreters}, Figure~\ref{fig:FIRST_images_weak_accreters}), while the remaining outflowing LERGs are extended radio sources. This is in contrast to the parent sample of 802 LERGs where we see $87\%$ of the population are compact radio sources, with the remainder being extended radio sources. If we only consider class~2 objects as extended, the fraction of compact sources rises to $\sim 98\%$. To summarize, although we find that about $50\%$ of the outflowing LERG populations are compact radio sources, it agrees with compact radio sources dominating the LERG population at $z<0.4$ where they account for $> 87\%$ in a large sample of LERGs.

    \begin{figure}[t]
        \centering
        \includegraphics[width=\columnwidth]{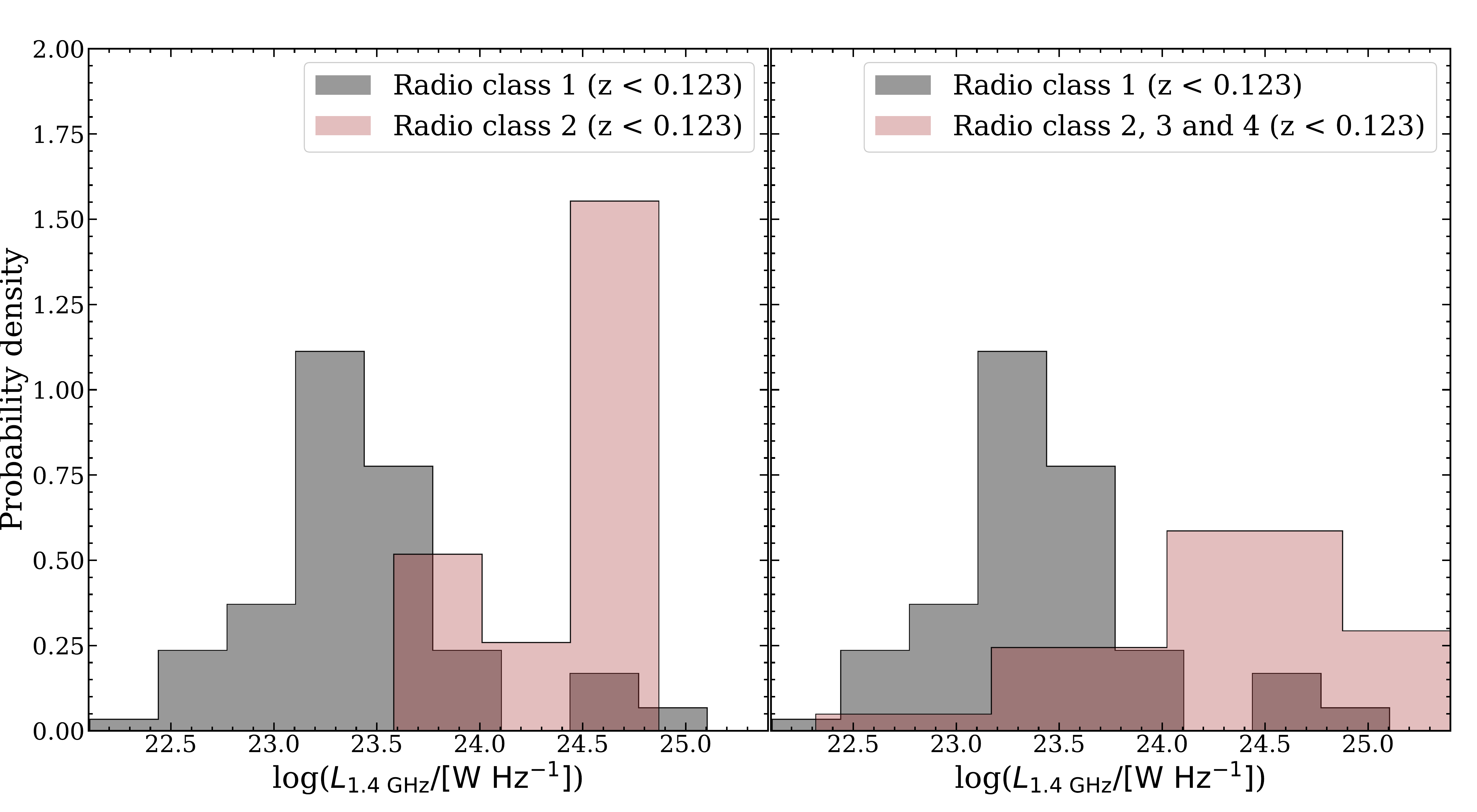}
        \includegraphics[width=\columnwidth]{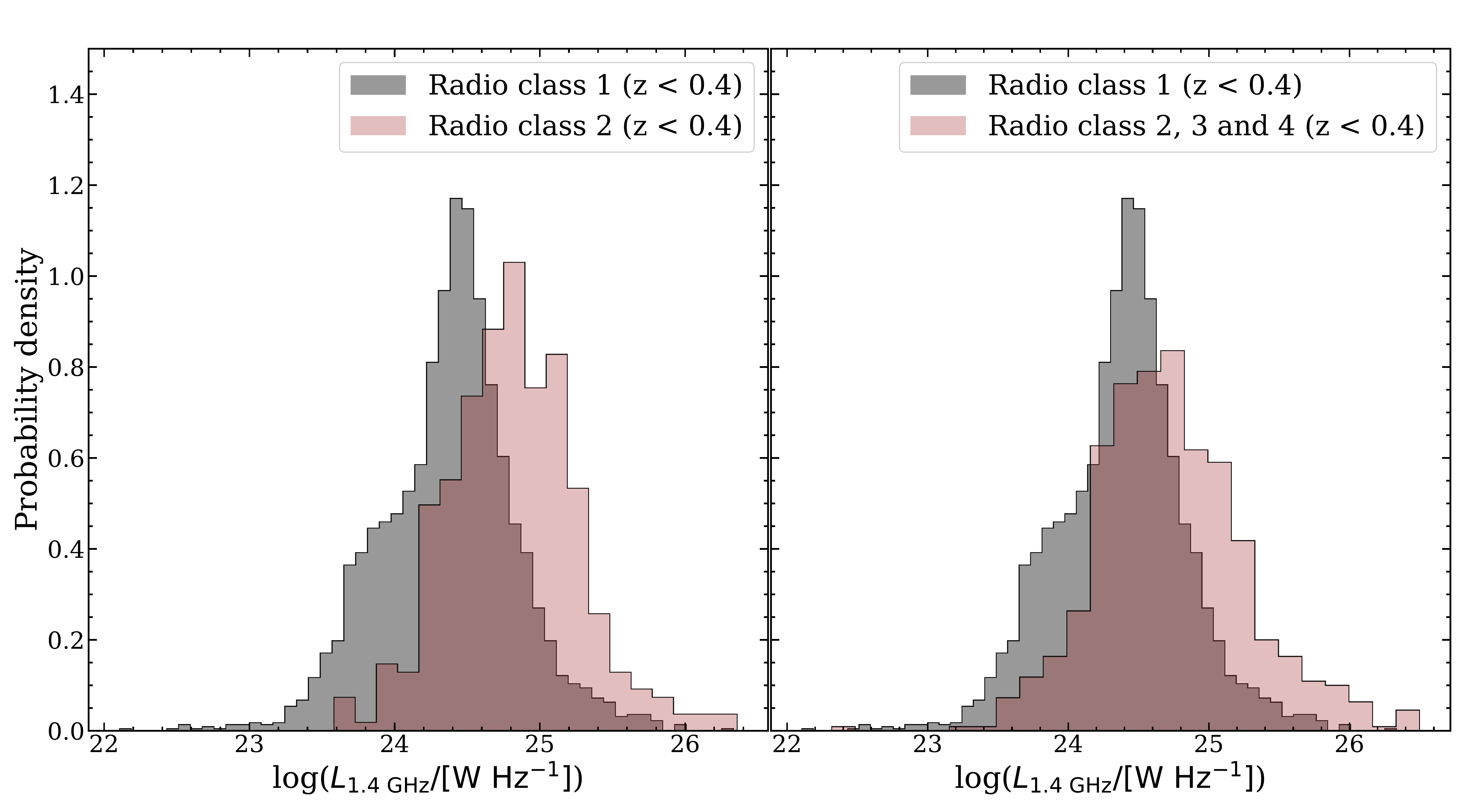}
        \caption{Normalized histograms of the $1.4~\text{GHz}$ radio-luminosity for the non-LERG radio AGN from \citet{best2012}. The area under each of these histograms is unity. The bin sizes are the nearest integer of the square root of the number of data points. \textit{Upper panels}: HERGs with $z < 0.123$. \textit{Lower panels}: HERGs with $z < 0.4$.}
        \label{fig:hist_Lrad_HERG}
    \end{figure}

\citet{mullaney2013} excluded LINERs in their analysis as LINERs tend to be objects with weak emission lines. It is possible that a sub-population of LINERs could represent a radio quiet analogue of LERGs. The sample of AGN analyzed by \citet{mullaney2013} only included type~1 and 2 Seyfert AGN, with HERGs being their closest counterparts amongst radio AGN. Therefore, we primarily focus on the radio morphology of HERGs from the sample of \citet{best2012}. In Figure~\ref{fig:hist_Lrad_HERG}, we plot histograms of the radio luminosities for the HERG population for $z < 0.123$ (upper panels) and $z < 0.4$ (the limit of the \citealt{mullaney2013} sample; lower panels). We notice the prevalence of compact radio sources until $z < 0.4$ relative to their extended counterparts (likely FR~I/II). In Figure~\ref{fig:hist_Lrad_HERG} the compact sources represent $91\%$, $79\%$, $88\%$, and $81\%$ of the HERG population, respectively, moving left to right, top to bottom. We further perform a two sample Kolmogorov-Smirnov test to check whether the compact and extended HERGs from Figure~\ref{fig:hist_Lrad_HERG} have similar distributions or not. We find the $p$-value $\ll 10^{-5}$ in each case indicating the compact and extended HERGs have very different distributions. In summary, we find that the compact radio sources account for $>$ 80\% of the HERG population at $z<0.4$. This is very similar to what we see for LERGs.

    \begin{figure}[t]
        \centering
        \includegraphics[width=\columnwidth]{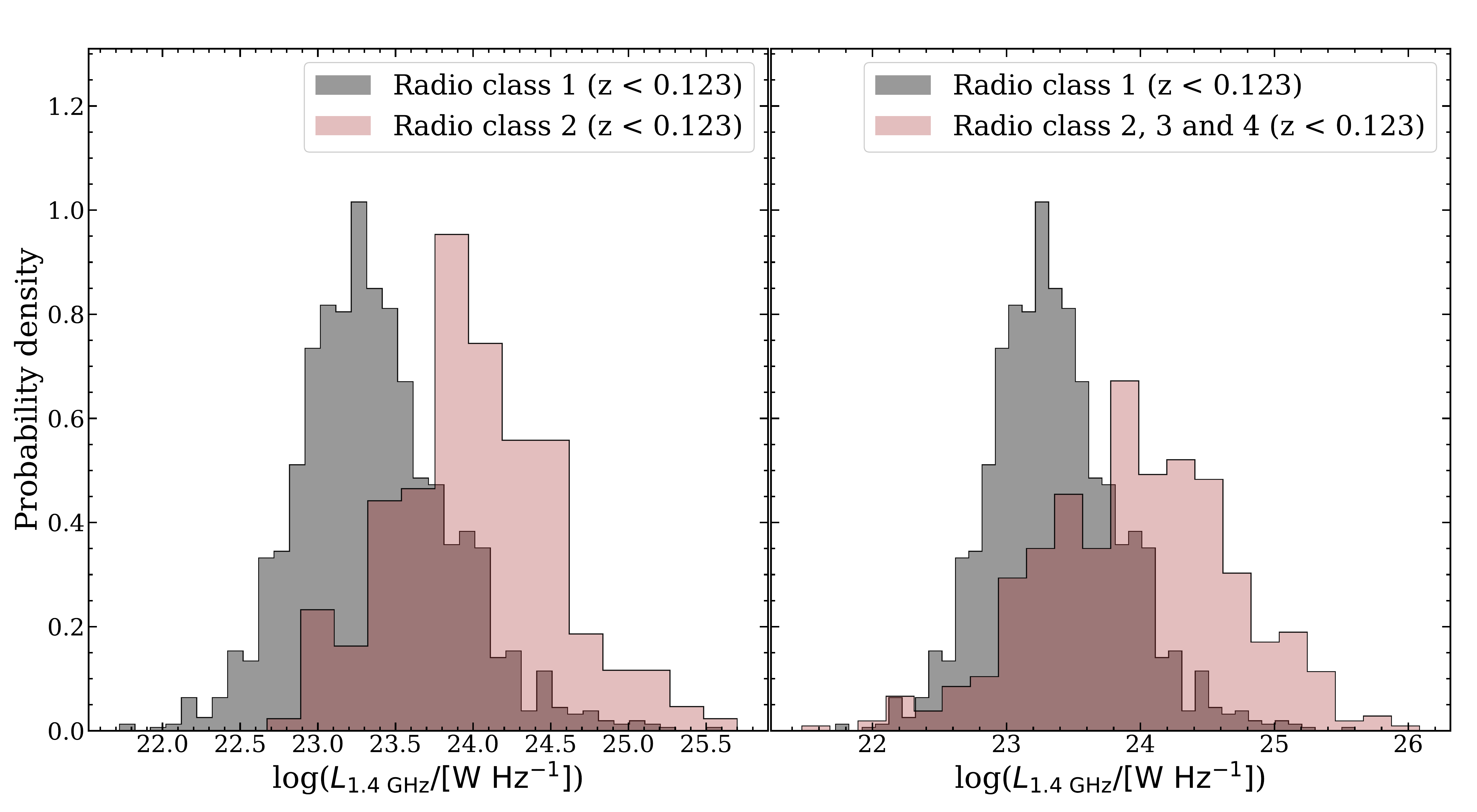}
        \includegraphics[width=\columnwidth]{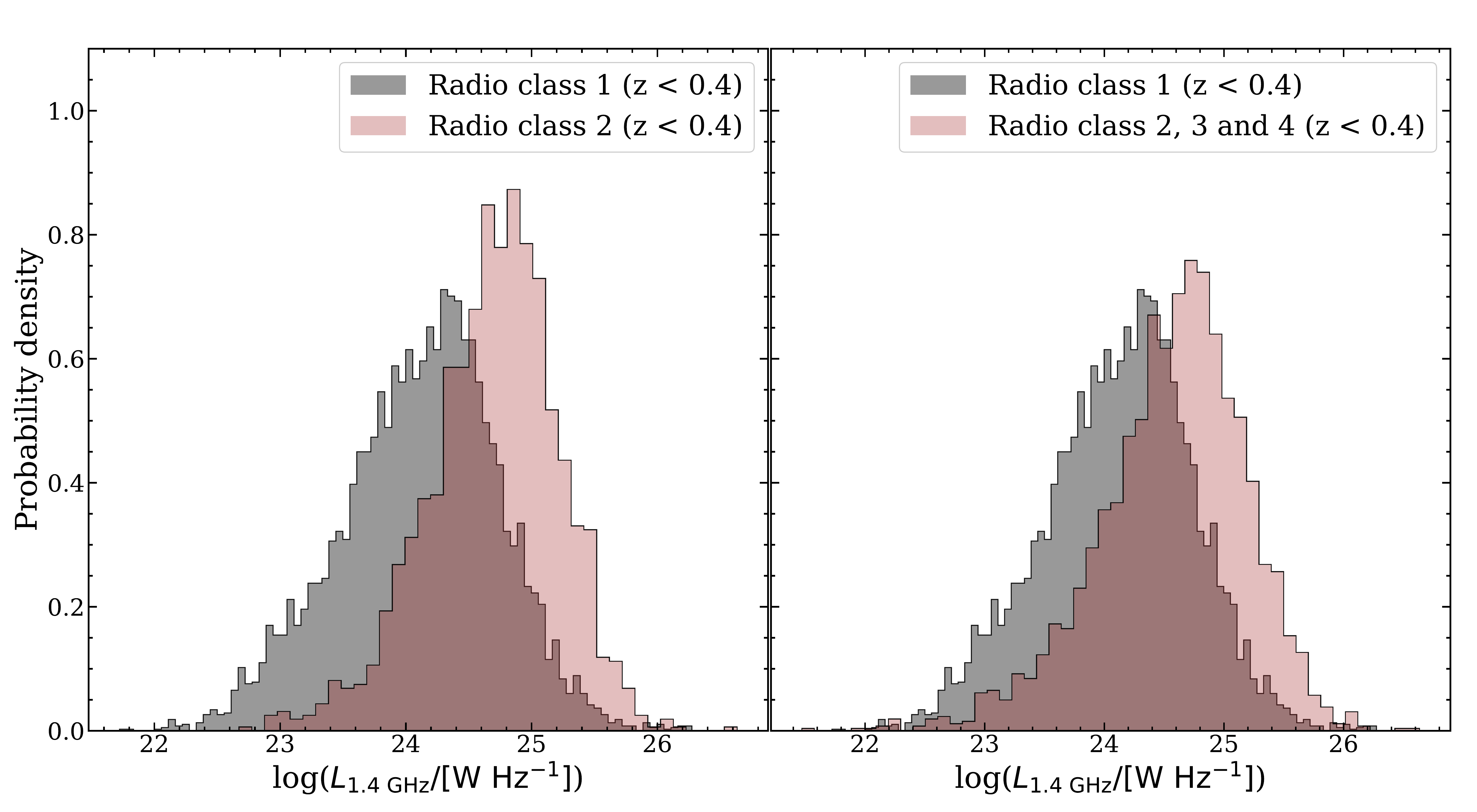}
        \caption{Normalized histograms of the $1.4~\text{GHz}$ radio-luminosity for the LERG radio AGN from \citet{best2012}, similar to Figure~\ref{fig:hist_Lrad_HERG}.}
        \label{fig:hist_Lrad_LERG}
    \end{figure}

As a next step, we study the radio morphologies of the LERG population by investigating the fraction of compact sources as a function of radio power in LERGs. In Figure~\ref{fig:hist_Lrad_LERG}, we plot the histograms of the radio luminosities for the LERG population for $z < 0.123$ (upper panels) and $z < 0.4$ (lower panels). We again notice the prevalence of compact radio sources until $z < 0.4$ relative to their extended counterparts (likely FR~I/II). In Figure~\ref{fig:hist_Lrad_LERG} the compact sources represent $89\%$, $76\%$, $81\%$, and $73\%$, of the LERG population, respectively, when moving from left to right, top to bottom, with an average $> 80\%$. \citet{miraghaei2017} studied 227 LERGs at $0.03 < z< 0.1$ and found 92 FR~I ($40\%$), 32 FR~II ($15\%$) and 103 FR~0 ($45\%$), showing the high number of compact sources in the sample; this is basically a redshift effect (where more compact sources are found at lower redshifts). Additionally, a two sample Kolmogorov-Smirnov test with an asymptotic distribution reveals that the compact and extended LERGs also have different distributions ($p$-value $\ll 10^{-15}$).

    \begin{figure*}[t]
        \centering
        \includegraphics[width=\textwidth]{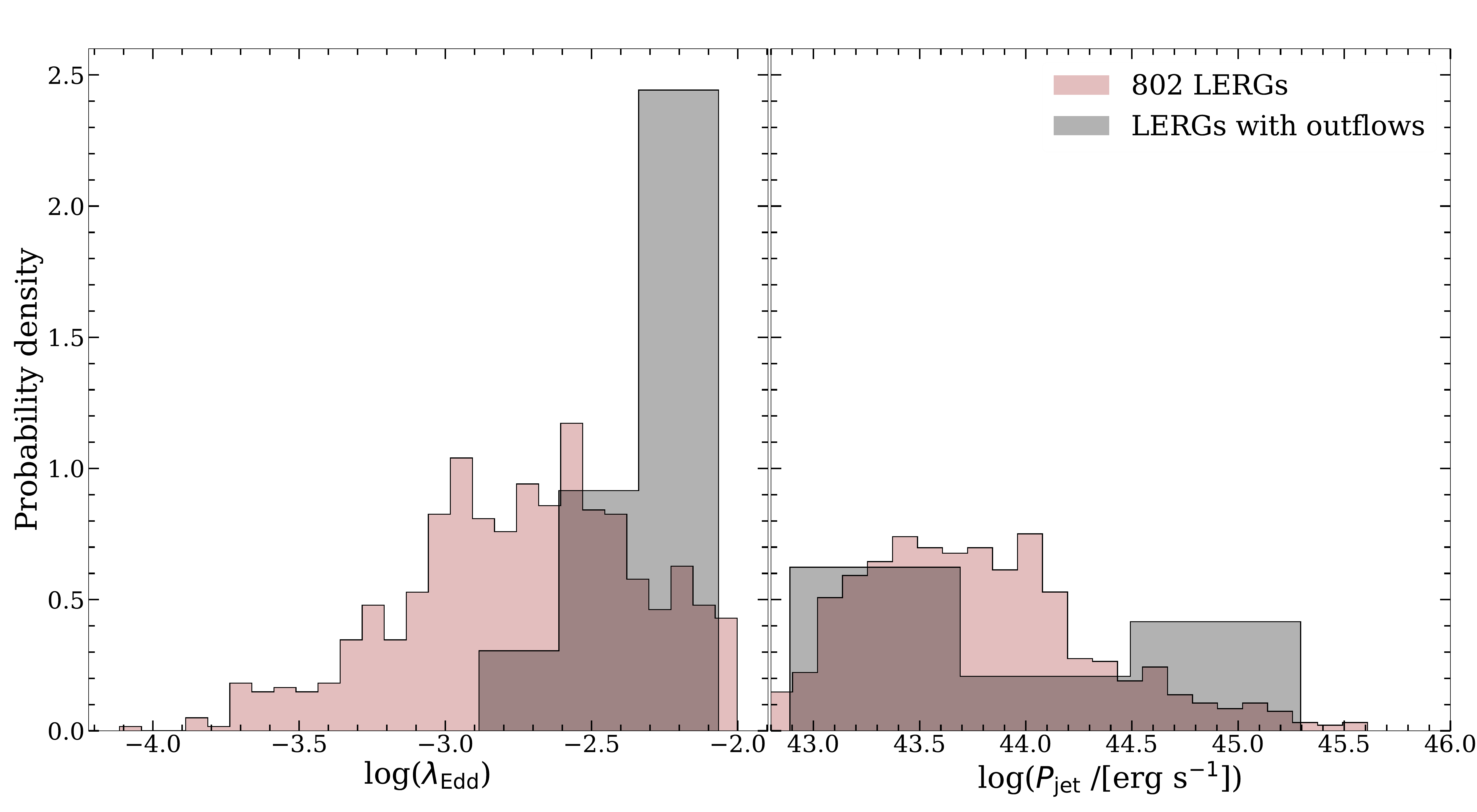}
        \caption{Normalized histograms of the 12 LERGs with outflows and their parent sample of 802 LERGs. The bin sizes are the nearest integer of the square root of the number of data points. \textit{Left}: Distribution of $\lambda_{\text{Edd}}$, restricted to $\lambda_{\text{Edd}} < 0.01$. \textit{Right}: Distribution of the jet power $P_{\text{jet}}$.}
        \label{fig:hist_lambdaEdd_Pjet}
    \end{figure*}
    
We also examine the radio morphology of the outflowing LERG population. The parent sample consisting of 802 LERGs, where 700 are radio class~1, 31 are class~2, 17 are class~3, and 54 are class~4. Therefore, 700 LERGs could be compact radio sources whereas 102 LERGs are extended radio sources. As for the outflowing LERGs, seven are radio class~1, two are class~3, and three are class~4, implying that we have seven compact radio sources and five extended radio sources. This is very similar to the visual inspection classification we performed. As a result, we see only $1\%$ of the compact radio sources show outflows whereas $\approx 20\%$ of extended radio sources show outflows in LERGs. One caveat is the small sample (12) of LERGs with outflows investigated here. It is therefore difficult to reach any firm conclusion given this modest sample size.

On average, we see that the compact radio sources are more than $80\%$ of the local AGN population up to $z < 0.4$. \citet{mullaney2013} stated that the sources with multiple FIRST matches, $> 80\%$ of the AGN in the radio luminosity range $3 \times 10^{23} < L_{1.4~\text{GHz}} < 3 \times 10^{24}~\text{W Hz}^{-1}$, could be radio cores. We find that the population of compact radio AGN in HERGs and LERGs peaks between $10^{23} - 10^{25}~\text{W Hz}^{-1}$ at both $z < 0.123$ and $z < 0.4$. This is the same radio luminosity range where the $FWHM_{\text{avg}}$ peaks in the work by \citet{mullaney2013}.

\citet{mullaney2013} noticed the peak $FWHM_{\text{avg}}$ in the region $10^{23} < L_{1.4~\text{GHz}} < 10^{25}~\text{W Hz}^{-1}$ which is highly dominated by compact radio AGN as shown above. Therefore if we analyze the radio AGN in that region, we will be mostly looking at the compact radio AGN. We observe that the [\ion{O}{3}] line-width does not peak within the region $10^{23} < L_{1.4~\text{GHz}} < 10^{25}~\text{W Hz}^{-1}$ for the LERGs with ionized outflows, and the population of compact radio sources has a peak within $10^{23} < L_{1.4~\text{GHz}} < 10^{25}~\text{W Hz}^{-1}$ in a sample of low-excitation and high-excitation radio AGN. Thus, the preference for outflows in compact sources found by \citet{mullaney2013} may simply be due to the fact that compact sources dominate the radio source population at $z < 0.4$.

\subsection{Are ionized outflows related to accretion disks or radio jets?}\label{subsec:disk_or_jet}

Previously the outflows in LERGs were usually connected to the acceleration from the bow shock of the radio jet \citep[e.g.,][]{morganti2005b,emonts2005,labiano2013,schulz2018}. However, \citet{woo2016} analyzed a sample of $\sim$ 24,000 type~2 AGN up to $z < 0.3$, and concluded that the outflows in AGN could be linked to AGN accretion rather than their radio luminosity. The AGN in the sample of \citet{woo2016} consisted of both LINERs and Seyferts, and the outflows were strongly correlated with [\ion{O}{3}] luminosity and Eddington ratio. Some of those LINER AGN could possibly represent a radio-quiet analogue of the LERGs as they tend to have weak emission lines. Therefore, based on our observations, we investigate the key driving force behind these outflows in LERGs. Our key results are as follows:
\begin{itemize}
    \item We see that only $\sim 1.5\%$ of the LERGs in our sample show signs of outflows in [\ion{O}{3}]. We used the selection criterion mentioned in Section~\ref{subsec:[OIII]_model} to determine the fraction of the type~2 optically selected AGN with ionized outflows in the ALPAKA catalogue \citep{mullaney2013}. We find that $\sim 35\%$ of the optically selected type~2 AGN show signs of ionized outflows. \citet{woo2016} reported that $\sim 45\%$ of type~2 AGN at $z < 0.3$ show signs of outflows in [\ion{O}{3}]. This indicates that having a radio jet in a low excitation AGN is neither necessary nor sufficient for generating an ionized outflow.
    
    \item FR~I jets are thought to interact strongly with their ambient media \citep[e.g.,][]{laing2014} and might be the best candidates for driving a gaseous outflow. However, we find ionized outflows in LERGs with a range of radio morphologies: compact (FR~0), FR~I, and FR~II. Therefore there is no preference for any type of radio morphology in order to drive ionized outflows.
    
    \item We calculate the jet power $P_{\text{jet}}$ using the relation given by \citet{cavagnolo2010} and present the distribution of jet powers in the right panel of Figure~\ref{fig:hist_lambdaEdd_Pjet}. We do not observe any noticeable trend between $P_{\text{jet}}$ and the occurrence of the outflows. We perform a two sample Kolmogorov-Smirnov test, and report a $p$-value of $0.33$, which reveals that the jet powers in the outflowing LERGs could have similar distributions as their parent sample of 802 LERGs.
    
    \item We present the distribution of $\lambda_{\text{Edd}}$ in the left panel of Figure~\ref{fig:hist_lambdaEdd_Pjet}. We see a sharp increase in the number of LERGs exhibiting [\ion{O}{3}] outflows as all of the 12 LERGs showing ionized outflows have $\lambda_{\text{Edd}} > 0.001$ in the left panel. As $\lambda_{\text{Edd}}$ increases, the number of LERGs with outflows increases. The largest number (8 out of 12) of LERGs with outflows are near $\lambda_{\text{Edd}} \sim 0.01$. We also perform a two sample Kolmogorov-Smirnov test, and report a $p$-value of $\sim 10^{-4}$, which reveals that the Eddington ratios in the outflowing LERGs are distributed very differently compared to their parent sample of 802 LERGs.
\end{itemize}

These findings suggest that the launching mechanism of the ionized outflows could be related to radiation pressure from the accretion disk, rather than the radio jet.

\citet{blandford1999} proposed the idea of an advection-dominated inflow-outflow solution, in which a small fraction of gas accreted onto the SMBH provides the energy to launch outflows. They also stated that the outflows could be self-collimating and form jets. One alternative is that before the accreted material reaches the inner accretion disk, the radiation pressure from accretion disk photons pumps the material outwards. As the ejected material travels farther, it further gets accelerated outwards from the shock of the radio jet \citep{capetti1999,tadhunter2001}. As the material crosses kpc scales carried out by the shock of the jet, it is located on the jet's axis. This could possibly explain why \citet{emonts2005,labiano2013,schulz2018} observed the outflowing atomic and ionized gas on the radio jet's axis.

\subsection{Feedback from LERGs}\label{subsec:LERG_feedback}

AGN feedback has been a widely debated topic - as it is a key ingredient to numerous semi-analytical models and numerical simulations, dealing with galaxy evolution. As pointed out by numerous studies  \citep[e.g.,][]{dimatteo2005,bower2006,croton2006,somerville2008,schaye2015}, AGN feedback is essential to quench star formation at larger stellar masses and to reproduce the host galaxy properties. In these models a fraction of the gas escapes from the host galaxy's potential due to galaxy wide outflows \citep{harrison2014}. This suppresses future star formation, regulates the central SMBH growth, and enriches the larger scale environment with metals \citep{harrison2014}. In this section we will discuss whether the ionized outflows in LERGs are reaching galaxy-wide scales, and more importantly, if they have an impact on the evolution of their host galaxies or not.

In order to influence the host galaxy's evolution, these outflows must necessarily be large enough to expel gas out of the entirety of the host galaxy. In recent years numerous studies have found that outflows from luminous quasars ($10^{41} < L_{\text{[O III]}} < 10^{44}~\text{erg s}^{-1}$) could reach up to $\sim 10~\text{kpc}$ scales \citep[e.g.,][]{genzel2014}.
We compare outflow sizes against [\ion{O}{3}] luminosity, $L_{\text{[O III]}}$, for a selection of luminous quasars as derived by \citet{lonsdale2003,davies2004,nesvadba2006,nesvadba2008,reyes2008,veilleux2009,harrison2012,rupke2013,liu2013a,liu2013b,harrison2014,genzel2014,cresci2015,carniani2015,brusa2015a,brusa2015b,perna2015,brusa2016,wylezalek2016,bischetti2017,duras2017}. We notice that the highly luminous AGN usually have higher outflow sizes, with the exception of the sample of \citet{carniani2015} where the outflow sizes are $\sim 0.5-2~\text{kpc}$. In the sample of \citet{genzel2014}, as the AGN become more luminous the outflow sizes become constant at $\sim 1.5~\text{kpc}$. A similar cutoff of outflow sizes is seen in the studies by \citet{bischetti2017,duras2017} where the outflow sizes reach a constant value of $\sim 8~\text{kpc}$. For the other studies, we do notice an increasing trend between $L_{\text{[O III]}}$ and $R_{\text{out}}$. Consequently,we perform a Spearman's rank correlation analysis to check for a correlation between $L_{\text{[O III]}}$ and $R_{\text{out}}$, and report a $p$-value of $\sim 0.69$, concluding no correlation. Between $10^{42} < L_{\text{[O III]}} < 10^{44}~\text{erg s}^{-1}$, $R_{\text{out}}$ is in the range $\sim 2-20~\text{kpc}$. For our LERGs, $R_{\text{out}} \sim 1~\text{kpc}$ with one exception where $L_{\text{[O III]}} \sim 10^{42}~\text{erg s}^{-1}$ and $R_{\text{out}} \sim 2.5~\text{kpc}$. This indicates that the LERGs, which are low-luminosity sources, could possibly have outflows which extend to the central $1~\text{kpc}$ of their host galaxies.

\begin{figure*}[t]
        \centering
        \includegraphics[width=\textwidth]{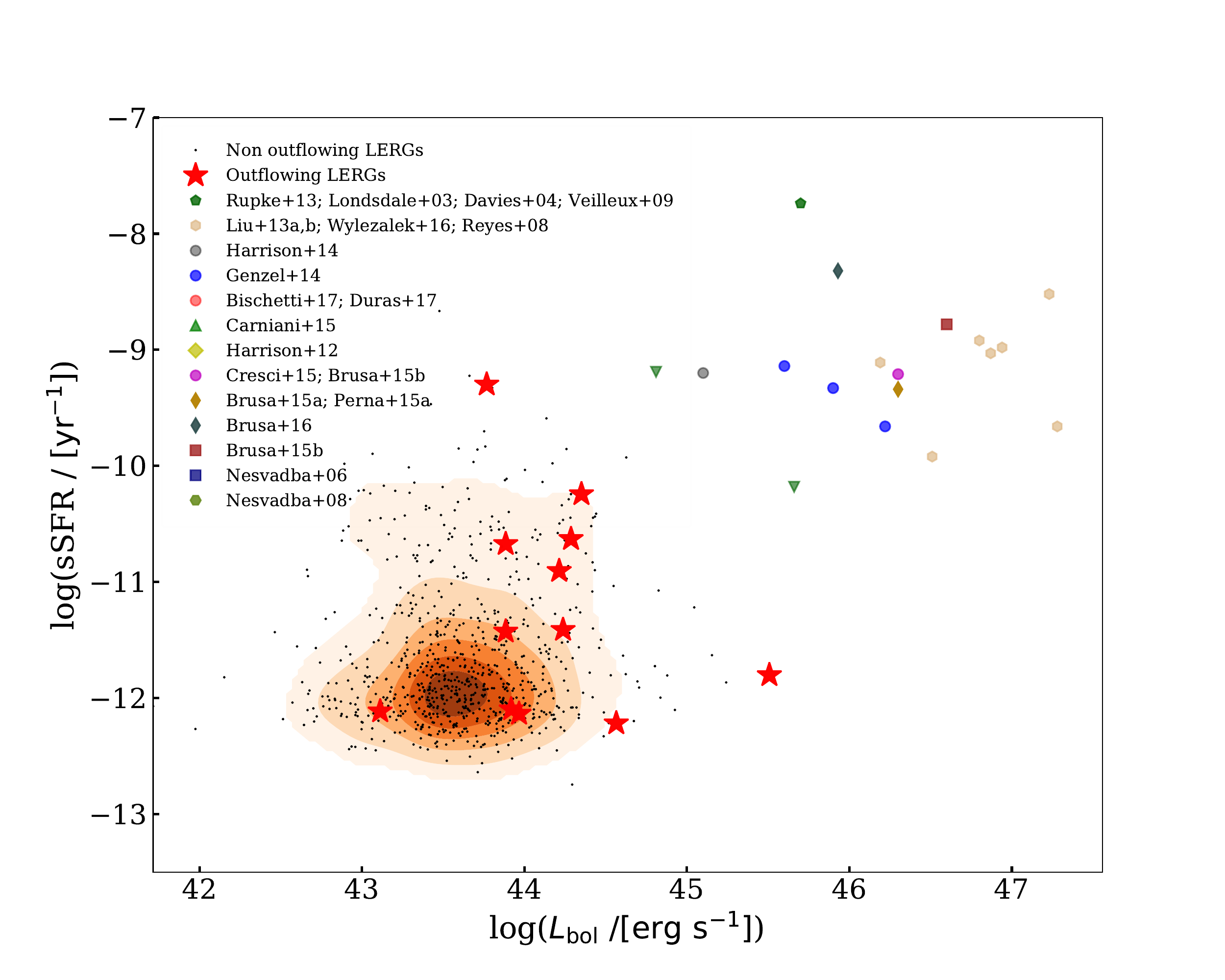}
        \caption{Specific star formation rate (sSFR) versus AGN bolometric luminosity ($L_{\text{bol}}$) for our outflowing LERGs (large red stars) along with luminous quasars from the literature \citep[other symbols,][]{lonsdale2003,davies2004,nesvadba2006,nesvadba2008,reyes2008,veilleux2009,rupke2013,liu2013a,liu2013b,harrison2012,harrison2014,genzel2014,cresci2015,carniani2015,brusa2015a,brusa2015b,perna2015,brusa2016,wylezalek2016,bischetti2017,duras2017}. Contours denote the distribution of non-outflowing LERGs (small black dots) in the $L_{\text{bol}}-\text{sSFR}$ plane.}
        \label{fig:sSFR_vs_Lbol}
    \end{figure*}

We now turn to analyze whether these LERG-driven outflows significantly quench star formation in their host galaxies or not. As outflows are of multi-phase gas \citep{harrison2014,husemann2016,husemann2019} and the majority of the mass is carried out by molecular outflows, it is difficult to constrain the complex nature of multi-phase gas. However, based on our observations of the ionized outflows within the central $1~\text{kpc}$ we will attempt to infer the effect of these outflows on the star formation within the host galaxies. We previously estimated a star formation rate of $1-12~M_{\sun}~\text{yr}^{-1}$ (see Table~\ref{tab:lerg_feedback}). Only one object, {J115141.05+334144.5}, has a $\text{SFR} > 232~M_{\sun}~\text{yr}^{-1}$. If the host galaxy has a large stellar mass then it is possible it could automatically have a high SFR due to an abundance of cold gas. To mitigate this issue, we investigate the specific star formation rate, $\text{sSFR} = \text{SFR}/M_{*}$ \citep{behroozi2013,schaye2015}, where we adopt the stellar masses from the MPA/JHU catalogue. In Figure~\ref{fig:sSFR_vs_Lbol}, we plot the sSFR against the AGN bolometric luminosity, $L_{\text{bol}}$, for our LERGs along with the aforementioned studies of luminous quasars. We additionally show the locations of the other 790 LERGs without outflows as points with corresponding contours. Figure~\ref{fig:sSFR_vs_Lbol} shows that the LERGs with outflows tend to have higher $L_{\text{bol}}$ than the LERGs without outflows. We notice that for our outflowing LERGs the sSFR is in the range $10^{-12} < \text{sSFR} < 10^{-9}~\text{yr}^{-1}$. We see the similar range of sSFR for the non-outflowing LERGs, with a concentration of $L_{\text{bol}}$ values between $10^{43} < L_{\text{bol}} < 10^{44}~\text{erg s}^{-1}$. For the luminous quasars from the previous studies, the sSFR is in the range $10^{-10} < \text{sSFR} \lesssim 10^{-7.5}~\text{yr}^{-1}$. This discrepancy indicates that the LERGs are a different class of object and have much lower sSFR possibly due to gas depletion.

\begin{figure*}[t]
        \centering
        \includegraphics[width=\textwidth]{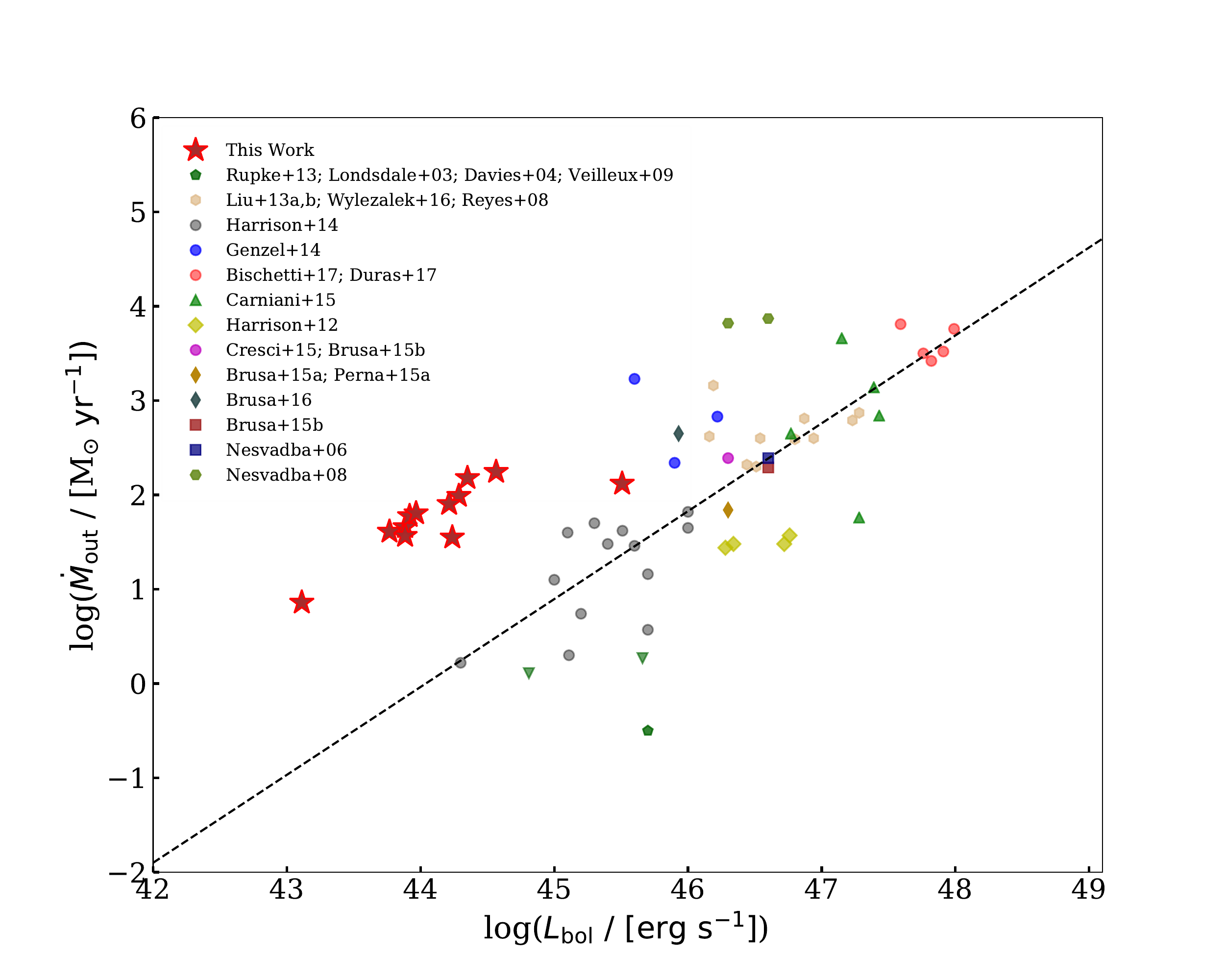}
        \caption{Mass outflow rate ($\dot{M}_{\text{out}}$) versus AGN bolometric luminosity ($L_{\text{bol}}$) for our outflowing LERGs along with luminous quasars from the literature. Symbols are as in Figure~\ref{fig:sSFR_vs_Lbol}. The straight line highlights the increasing trend between $L_{\text{bol}}$ and $M_{\text{out}}$ for the luminous quasars. The outflowing LERGs lie above the straight line.}
        \label{fig:Mout_vs_Lbol}
    \end{figure*}

Despite the low star formation rates observed in the LERGs, the outflow needs to expel significant gas mass out of the galaxy in order to quench star formation and to produce the observable host galaxy properties. We estimate mass outflow rates of $7-150~M_{\sun}~\text{yr}^{-1}$ (see Table~\ref{tab:lerg_feedback}) using Equation~\ref{eq:M_out}. Comparing it with the SFR, we notice that they are $\approx 3 –- 60$ times greater than the SFRs of the outflowing LERGs, with one exception at $0.5$ times for {J115141.05+334144.5} where we observe a prominent red wing. These mass loading factors, $\eta$, are often comparable to that of the luminous quasars from the samples of \citet{liu2013a,liu2013b,genzel2014,bischetti2017,duras2017}, but much higher than the sample of \citet{harrison2012,harrison2014} who traced the ionized phase of the outflows, and other works looking at the different gas phases \citep[e.g.][]{martin1999,heckman2000,newman2012,cicone2014}. In the sample of \citet{carniani2015}, we see a diversity in mass loading factors. The quasars at $L_{\text{bol}} < 10^{46}~\text{erg s}^{-1}$ have $\eta <1$, whereas two quasars from the sample have $\eta > 20$ at $L_{\text{bol}} \sim 10^{47}~\text{erg s}^{-1}$ \citep{carniani2015}. Strikingly, one quasar at $L_{\text{bol}} \sim 10^{47}~\text{erg s}^{-1}$ in their sample has $\eta < 1$. In Figure~\ref{fig:Mout_vs_Lbol}, we plot the mass outflow rates, $\dot{M}_{\text{out}}$, against $L_{\text{bol}}$ for our LERGs along with the results from the previously mentioned studies of luminous quasars. We notice an increasing trend between $L_{\text{bol}}$ and $\dot{M}_{\text{out}}$ for both our LERGs and the quasars. The most striking feature is that the LERGs typically have higher mass outflow rates per $L_{\text{bol}}$ than the other AGN. This could be an effect due to the existence of the radio source and how the bow shock can accelerate gas clouds \citep[e.g.,][]{odea2002}. In addition,  jets may be able to entrain gas clouds \citep{deyoung1986,deyoung1993,fedorenko1996,laing2002,wang2009}, as the bow shock from the radio jet could sweep out material further out in the galaxy, resulting in a higher $\eta$.

\newpage
\section{Summary and Conclusions}\label{sec:summary_and_conclusions}

We have presented SDSS optical spectroscopic observations covering the [\ion{O}{3}]~$\lambda\lambda 4959, 5007$ emission lines of 802 LERGs with $0.01 < z < 0.3$. Our targets were selected from a parent sample of 18,286 radio galaxies with $z < 0.3$ from \citet{best2012} and we demonstrate the presence of outflows in the [\ion{O}{3}]~$\lambda\lambda 4959, 5007$ emission lines. In short our findings are as follows:
\begin{itemize}
    \item Only 12 out of 802 LERGs display visible broad wings in the [\ion{O}{3}]~$\lambda\lambda 4959, 5007$ emission lines. This represents $\sim 1.5\%$ of the parent sample (Section~\ref{subsec:[OIII]_model}).
    
    \item Six out of these 12 LERGs are compact radio sources, three are FR~I, and the remaining three are FR~II sources (Section~\ref{subsec:radio_morph}). The population of compact radio sources represents $\sim 80\%$ of the radio AGN population until $z < 0.4$. Thus there is no preference for outflows to be found in a given radio source morphology (Section~\ref{subsec:compact_sources}).
    
    \item The numbers of LERGs with outflows show no significant trend with jet power (Section~\ref{subsec:disk_or_jet}).
    
    \item The number of LERGs with outflows increases with increasing Eddington ratio. Eight out of the 12 outflowing LERGs have Eddington ratios close to $0.01$ (Section~\ref{subsec:disk_or_jet}).
    
    \item LERGs with outflows typically have $R_{\text{out}} \sim 1~\text{kpc}$, with only one source having $R_{\text{out}} \sim 2.5~\text{kpc}$. Although other luminous quasars have higher outflow sizes, there is no correlation between $L_{\text{[O III]}}$ and $R_{\text{out}}$ (Section~\ref{subsec:LERG_feedback}).
    
    \item LERGs with outflows have lower sSFR ($10^{-12} < \text{sSFR} < 10^{-9}~\text{yr}^{-1}$) than luminous quasars. This is indicative that LERGs are a different class of AGN (Section~\ref{subsec:LERG_feedback}).
    
    \item LERGs with outflows show higher mass outflow rates than other luminous quasars for a given $L_{\text{bol}}$. The existence of a radio source could possibly enhance the mass loading (Section~\ref{subsec:LERG_feedback}).
\end{itemize}

All these results indicate that the ionized outflows in LERGs may be linked to the accretion disk, not the radio jet.

\acknowledgments
We thank the anonymous referee for their thoughtful and helpful comments that improved the quality of the manuscript. The work of MS was supported in part by the University of Manitoba Faculty of Science Graduate Fellowship (Cangene Award), and by the University of Manitoba Graduate Enhancement of Tri-Council Stipends (GETS) program. The work of CPO, YAG, and SAB was supported by a grant from the Natural Sciences and Engineering Research Council of Canada (NSERC). The work of CLF was supported in part by the University of Manitoba Graduate Fellowship, and by the University of Manitoba GETS program.

Funding for the SDSS and SDSS-II has been provided by the Alfred P. Sloan Foundation, the Participating Institutions, the National Science Foundation, the U.S. Department of Energy, the National Aeronautics and Space Administration, the Japanese Monbukagakusho, the Max Planck Society, and the Higher Education Funding Council for England. The SDSS Web Site is \url{http://www.sdss.org/}. The SDSS is managed by the Astrophysical Research Consortium for the Participating Institutions. The Participating Institutions are the American Museum of Natural History, Astrophysical Institute Potsdam, University of Basel, University of Cambridge, Case Western Reserve University, University of Chicago, Drexel University, Fermilab, the Institute for Advanced Study, the Japan Participation Group, Johns Hopkins University, the Joint Institute for Nuclear Astrophysics, the Kavli Institute for Particle Astrophysics and Cosmology, the Korean Scientist Group, the Chinese Academy of Sciences (LAMOST), Los Alamos National Laboratory, the Max-Planck-Institute for Astronomy (MPIA), the Max-Planck-Institute for Astrophysics (MPA), New Mexico State University, Ohio State University, University of Pittsburgh, University of Portsmouth, Princeton University, the United States Naval Observatory, and the University of Washington.

The National Radio Astronomy Observatory is a facility of the National Science Foundation operated under cooperative agreement by Associated Universities, Inc.

This research has made use of NASA's Astrophysics Data System, as well as TOPCAT, an interactive graphical viewer and editor for tabular data \citep{taylor2005}, in addition to Astropy\footnote{\url{https://astropy.org}}, a community-developed core Python package for Astronomy \citep{astropy2013,astropy2018}.

This research has made use of the VizieR catalogue access tool\footnote{\url{http://vizier.u-strasbg.fr/viz-bin/VizieR}}, CDS, Strasbourg, France (DOI: 10.26093/cds/vizier). The original description of the VizieR service was published in \citet{ochsenbein2000}.

\facilities{Sloan, VLA}

\software{Astropy \citep{astropy2013,astropy2018}, Matplotlib \citep{hunter2007}, NumPy \citep{vanderwalt2011}, SciPy \citep{virtanen2020}, TOPCAT \citep{taylor2005}}

\end{document}